\UseRawInputEncoding
\documentclass[superscriptaddress, twocolumn, amsmath, amssymb, aps,pra, notitlepage,longbibliography]{revtex4-2} 
\usepackage{graphicx,graphics,epsfig,subfigure,times,bm,bbm,amssymb,amsmath,amsfonts,amsthm,mathrsfs,MnSymbol}
\usepackage{braket}
\usepackage[matrix,frame,arrow]{xypic}
\usepackage[normalem]{ulem}
\usepackage{slashed}
\usepackage{dcolumn}
\usepackage{tabularx}
\usepackage{amsopn}
\usepackage{bigints}
\usepackage{color}
\usepackage{siunitx}
\usepackage[usenames,dvipsnames,svgnames,table]{xcolor}
\usepackage[english]{babel}
\usepackage{verbatim}
\usepackage{mathtools}
\definecolor{darkblue}{rgb}{0.0,0.0,0.3}
\usepackage[colorlinks=true,
            linkcolor=red,
            urlcolor= darkblue,
            citecolor=blue]{hyperref}

\usepackage{cleveref}
\crefname{equation}{Eq.}{Eqs.}
\crefname{section}{Sec.}{Sec.}

\newtheorem*{proposition}{Proposition}

\newcommand{\tr}{\operatorname{tr}}
\newcommand{\re}{\operatorname{Re}}
\newcommand{\dd}{\mathrm{d}}
\newcommand{\proj}[1]{\left|#1\right\rangle\kern-2pt\left\langle#1\right|}
\newcommand{\jump}[2]{\left|#1\right\rangle\kern-2pt\left\langle#2\right|}
\newcommand{\bbra}[1]{\left\llangle#1\right|}

\newcommand{\bbracckett}[3]{\left\llangle#1\middle|#2\middle|#3\right\rrangle}

\begin{document}
\title{Full counting statistics of electron-photon hybrid systems: Joint statistics and fluctuation symmetry}
\author{Tianyi Xiao}
\affiliation{Department of Physics, Institute for Quantum Science and Technology, Shanghai Key Laboratory of High Temperature Superconductors, International Center of Quantum and Molecular Structures, Shanghai University, Shanghai, 200444, China}
\author{Junjie Liu}
\email{jj\_liu@shu.edu.cn}
\affiliation{Department of Physics, Institute for Quantum Science and Technology, Shanghai Key Laboratory of High Temperature Superconductors, International Center of Quantum and Molecular Structures, Shanghai University, Shanghai, 200444, China}

\begin{abstract}
Electron-photon hybrid systems serve as ideal light-matter interfaces with broad applications in quantum technologies. These systems are typically operated dynamically under nonequilibrium conditions, giving rise to coupled electronic and photonic currents. Understanding the joint fluctuation behavior of these currents is essential for assessing the performance of light-matter interfaces that rely on electron-photon correlations. Here, we investigate the full counting statistics of coupled electronic and photonic currents in an experimentally feasible hybrid system composed of a double quantum dot coupled to an optical cavity. We employ the framework of quantum Lindblad master equation which is augmented with both electronic and photonic counting fields to derive their joint cumulant generating function--a treatment that differs significantly from existing studies, which typically focus on either electron or photon statistics separately. We reveal that the ratio between photonic and electronic currents, as well as their variances, can deviate from an expected quadratic scaling law in the large electron-photon coupling regime. Furthermore, we demonstrate that conventional modelings of photonic dissipation channels in quantum master equations must be modified to ensure that the joint cumulant generating function satisfies the fluctuation symmetry enforced by the fluctuation theorem. Our results advance the understanding of joint fluctuation behaviors in electron-photon hybrid systems and may inform the design of efficient quantum light-matter interfaces.  
\end{abstract}

\date{\today}
\maketitle

\section{Introduction}
Quantum hybrid systems~\cite{Xiang.13.RMP,Clerk.20.NP,Burkard.20.NRP} integrate the complementary advantages of distinct quantum systems, enabling broad applications in quantum information processing, communication, sensing and beyond. These applications rely on the generation and control of quantum correlations across complementary degrees of freedom. Among the diverse range of hybrid architectures, light-matter hybrid systems~\cite{Clerk.20.NP,Burkard.20.NRP,Kockum.19.NRP,Gutzler.21.NRP,Rivera.20.NRP,Diaz.19.RMP,Aspelmeyer.14.RMP} have emerged as particularly promising platforms due to their advanced control toolkit, capacity to achieve strong coupling, and relative ease of fabrication. Of particular interest are electron-photon hybrid systems, which combine electronic and photonic components. In such systems, both electrons and photons can serve as information-carrying degrees of freedom--acting as stationary and flying qubits, respectively--in applications such as quantum networks~\cite{Duan.10.RMP,Azuma.23.RMP} and quantum computing~\cite{Lim.06.PRA,Jiang.07.PRA}.

To enable desired functionalities, electron-photon hybrid systems are typically driven under nonequilibrium conditions, giving rise to sustained electronic and photonic currents. At the nanoscale, these currents exhibit intrinsic fluctuations characterized by nonzero higher-order moments. A detailed characterization of these fluctuation--particularly their higher-order correlations--is essential for evaluating the performance of such complex systems~\cite{Schweigler.17.N}. In response to this need, numerous studies have been conducted to investigate the full counting statistics of nonequilibrium electron-photon hybrid systems~\cite{agarwalla_photon_2019,Dahan.21.S,Xu.13.PRB,Nian.23.PRB,Wang.19.AS,Zenelaj.22.PRB,Samuelsson.14.PRB,Yanagimoto.23.CP,van_den_berg_charge-photon_2019}. Despite the capability of electron-photon hybrid systems to reach the strong electron-photon coupling regime~\cite{Kockum.19.NRP,GuX.17.PR}, existing works have largely focused on either photon statistics~\cite{Xu.13.PRB,agarwalla_photon_2019,Nian.23.PRB,Dahan.21.S,Zenelaj.22.PRB,Wang.19.AS} or electron statistics~\cite{Samuelsson.14.PRB}, while the joint statistics of electrons and photons remains largely unexplored--with Ref.~\cite{van_den_berg_charge-photon_2019} being a notable exception. Understanding the joint statistics is crucial for elucidating electron-photon correlations, which ultimately determine the utility of light-matter interfaces. In this regard, several open questions remain: (i) How do electron and photon statistics relate to each other, especially in the large electron-photon coupling regime? Ref.~\cite{van_den_berg_charge-photon_2019} reported that the ratios between the averaged photonic and electronic currents, as well as their variances, all scale quadratically with the electron-photon coupling strength. However, the generality of this scaling remains unclear. (ii) Whether the joint statistics satisfies the fluctuation theorem~\cite{esposito_fluctuation_2007,esposito_nonequilibrium_2009,Jarzynski.11.AR,Seifert.12.RPP,Andrieux.09.NJP}? As a refinement of the second law of thermodynamics, the fluctuation theorem ensures thermodynamic consistency of fluctuations~\cite{Liu.21.PRE} as it naturally leads to fundamental relations such as the Onsager-Casimir reciprocity and fluctuation-dissipation theorems~\cite{Saito.08.PRB}.

Here, we address these questions by investigating the full counting statistics of a prototypical electron-photon hybrid system consisting of a double quantum dot embedded in an optical cavity~\cite{agarwalla_photon_2019,Samuelsson.14.PRB,Zenelaj.22.PRB}. Such hybrid systems have been experimentally realized~~\cite{Frey.12.PRL,Petersson.12.N,liu_photon_2014,kulkarni_cavity-coupled_2014,Stehlik.16.PRX,Liu.S.15,Hartke.18.PRL}. We utilize the framework of quantum Lindblad master equation dressed by both electronic and photonic counting fields to derive the joint cumulant generating function (CGF) of electrons and photons~\cite{esposito_nonequilibrium_2009,landi_current_2024}. By doing so, we are able to treat the statistics of electronic and photonic currents on equal footing. 

We begin by adopting a conventional quantum Lindblad master equation for this electron-photon hybrid system~\cite{agarwalla_photon_2019} to examine the behaviors of averaged currents, second-order fluctuations as well as their respective ratios. We find that the ratio of the average photonic to electronic current exhibits a quadratic scaling with the electron-photon coupling strength--but only in the weak coupling regime. As the electron-photon coupling strength increases, significant deviations from this quadratic behavior emerge, indicating that the results reported in Ref.~\cite{van_den_berg_charge-photon_2019} may be generalizable to other models under weak coupling conditions. In contrast, we observe that the second-order electron-photon cross
correlations can become negative in this system, contradicting a positive ratio between the variances proposed in Ref.~\cite{van_den_berg_charge-photon_2019} and certainly disobeying the same quadratic scaling suggested for the variance ratio in that work.

We then investigate whether the fluctuation theorem~\cite{esposito_fluctuation_2007,esposito_nonequilibrium_2009,Jarzynski.11.AR,Seifert.12.RPP,Andrieux.09.NJP} is satisfied in this system. Our analysis reveals that the joint CGF derived from the conventional quantum Lindblad master equation~\cite{agarwalla_photon_2019} fails to satisfy the fluctuation symmetry required by the fluctuation theorem. This discrepancy stems from the fact that the conventional treatment of photon dissipation channels implicitly assumes a zero-temperature photonic environment, accounting only for photon loss in the quantum master equation~\cite{agarwalla_photon_2019, Gullans.15.PRL, Aspelmeyer.14.RMP}. To restore the fluctuation symmetry for the joint electron-photon statistics, we modify the conventional quantum Lindblad master equation by further including a photon gain channel based on the detailed balance condition. This extension effectively models the photonic environment at finite temperatures--a treatment we argue is essential for consistency, given that electronic reservoirs are already described at finite temperatures. Using this modified framework, we demonstrate that the resulting joint CGF satisfies the fluctuation symmetry, thus underscoring the necessity of treating the photonic bath as a finite-temperature reservoir in order to ensure thermodynamic consistency of quantum Lindblad master equation for such electron-photon hybrid systems. Finally, we reassess the ratios between the average photonic and electronic currents using the modified CGF and confirm that our earlier conclusions--originally drawn from the conventional quantum Lindblad master equation--remain qualitatively valid at low temperatures. This suggests that although the conventional quantum Lindblad master equation fails to satisfy the fluctuation symmetry, it nevertheless provides a good approximation to the cumulant behaviors of the electron-photon hybrid system at low temperatures.

The paper is structured as follows. In \Cref{sec:1}, we introduce the Hamiltonian of the electron-photon hybrid system and present the conventional quantum Lindblad master equation governing the dynamics of the coupled system. We also describe the incorporation of counting fields into the master equation and provide the corresponding definition of the CGF. In \Cref{sec:2}, we investigate the relationship between electron and photon statistics, focusing on the average currents, second-order correlation functions, and their respective ratios. In \Cref{sec:3}, we analyze the validity of the fluctuation symmetry in detail. Specifically, we propose a modified quantum Lindblad master equation that restores the fluctuation symmetry for the joint statistics. Using the resulting CGF--which now satisfies the fluctuation system--we reassess the ratio between the average photonic and electronic currents. We summarize our findings in \Cref{sec:4}. For the sake of clarity, we relegate lengthy analytical derivations and discussions to Appendices \ref{sec:convergence_of_scgf} and \ref{sec:condition_for_fluctuation_theorem}, which address the convergence of the CGF in the steady-state limit and the conditions for the CGF to satisfy the fluctuation symmetry, respectively. 

\section{Model and full counting statistics}\label{sec:1}
We start by detailing the electron-photon hybrid system under study and the full counting statistics framework based on quantum master equations. The electron-photon hybrid system consists of a double quantum dot (DQD) system coupled to an optical cavity. The state of a single electron in the DQD is described within a two-dimensional Hilbert space spanned by the basis vectors \(\ket{L}\) and \(\ket{R}\), representing occupation of the left and right dot, respectively. To account for the possibility of an empty DQD, we introduce a zero-particle state \(\ket{0}\), resulting in a total Hilbert space of dimension three. The Hamiltonian of the DQD, which includes the tunneling of electrons between two dots, is given by
\begin{equation}
\hat H_{\text{DQD}} = \frac{\epsilon}{2} \left(\proj{L} - \proj{R}\right) + \tau \left( \jump{R}{L} + \jump{L}{R} \right).
\end{equation}
Here, \(\epsilon\) is the energy difference between the two dots and \(\tau\) describes the tunneling strength.

Each dot is connected to an electronic reservoir with a Hamiltonian reading (\(\alpha \in \{L, R\}\))
\begin{equation}
\hat H_\alpha = \sum_{k} \mathcal{E}_{k\alpha} \hat c_{k\alpha}^\dagger \hat c_{k\alpha},
\end{equation}
where the tunneling interaction between the DQD and the lead $\alpha$ is given by
\begin{equation}
\hat H_{\text{DQD}-\alpha} = \sum_k S_{k\alpha}\hat c_{k\alpha}\jump{\alpha}{0} + \text{H.c.}.
\end{equation}
Here, \(\mathcal{E}_{k\alpha}\) is the energy of $k$th mode in lead $\alpha$ with a corresponding annihilation operator \(\hat c_{k\alpha}\). \(S_{k\alpha}\) is the coupling strength between the lead mode and the DQD. ``H.c." is short for Hermitian conjugate.

As for the cavity part, we consider a single-mode cavity described by
\begin{equation}
\hat H_{\text{cavity}} = \omega \hat a^\dagger \hat a,
\end{equation}
where \(\omega\) and \(\hat a\) are the frequency and the annihilation operator of the cavity mode, respectively.
The cavity is further coupled to its own photonic reservoir which can be a transmission line in experiments. The Hamiltonian of the photonic reservoir is given by
\begin{equation}
\hat H_{\gamma} = \sum_{k} \omega_{k} \hat a_{k}^\dagger \hat a_{k},
\end{equation}
where \(\omega_{k}\) is the frequency and \(\hat a_{k}\) is the annihilation operator of the \(k\)-th mode in the photonic reservoir.
The tunneling between the cavity and the reservoir is given by
\begin{equation}
\hat H_{\text{cavity}-\gamma} = \sum_k \xi_{k} \hat a_{k}^\dagger \hat a + \text{H.c.}.
\end{equation}
Here \(\xi_{k}\) is the coupling strength between the \(k\)-th reservoir mode and the cavity mode. The interaction between the DQD and the cavity mode is given by a dipole interaction described by the Hamiltonian
\begin{equation}
\hat H_{\text{DQD}-\text{cavity}} = g\left( \jump{R}{L} + \jump{L}{R} \right)\left(\hat a^\dagger + \hat a\right),
\end{equation}
where \(g\) is the coupling strength between the DQD and the cavity mode.

\begin{table*}
\caption{Details of dissipation channels in \Cref{eq:master_equation} and our conventions in channel indices, current number and counting weights.\label{tab:channel_numbering}}
	\begin{tabular}{|c|c|c|c|c|c|c|c|c|c|}
		\hline
		Channel Number \(j\) & 1 & 2 & 3 & 4 & 5 & 6 & 7 & 8 & 9 \\
		\hline
		Related reservoir & \multicolumn{4}{c|}{Left electron lead} & \multicolumn{4}{c|}{Right electron lead} & Photon reservoir \\
		\hline
		Jump Operator (coefficient omitted) & \(\jump{0}{g}\) & \(\jump{g}{0}\) & \(\jump{0}{e}\) & \(\jump{e}{0}\)  & \(\jump{0}{g}\) & \(\jump{g}{0}\) & \(\jump{0}{e}\) & \(\jump{e}{0}\) & \(\hat a\) \\
		\hline
		Current number \(\zeta_j\) & \multicolumn{4}{c|}{\(1\)} & \multicolumn{4}{c|}{Not counted} & \multicolumn{1}{c|}{\(2\)} \\
		\hline
		Counting weight \(\nu_j\) & \(-1\) & \(1\) & \(-1\) & \(1\) & \multicolumn{4}{c|}{Not counted} & \(1\) \\
		\hline
	\end{tabular}
\end{table*}

By diagonalizing \(\hat H_{\text{DQD}}\), we obtain its eigenstates as
\begin{equation}
\begin{aligned}
	\ket g&=\cos\left(\frac{\theta}{2}\right)\ket{R}-\sin\left(\frac{\theta}{2}\right)\ket{L},\\
	\ket e&=\sin\left(\frac{\theta}{2}\right)\ket{R}+\cos\left(\frac{\theta}{2} \right)\ket{L}.
\end{aligned}
\end{equation}
Here, \(\epsilon_g = -\Omega/2\) and \(\epsilon_e = \Omega/2\) are the corresponding eigenvalues, respectively. \(\Omega = \sqrt{\epsilon^2 + 4\tau^2}\) is the energy difference, and the mixing angle is defined by \(\theta = \arctan(2\tau/\epsilon)\). We note that the zero-particle state \(\ket{0}\) remains unaltered by this transformation. In the following, we will work exclusively in this eigenbasis $\{|0\rangle,|g\rangle,|e\rangle\}$ for the DQD system.

Under weak system-reservoir coupling conditions, the dynamics of the composite DQD-cavity system can be described by quantum master equations. At this moment, we adopt an established quantum Lindblad master equation for this system~\cite{agarwalla_photon_2019} 
\begin{equation}\label{eq:master_equation}
\frac{d\hat\rho}{dt} = -i\left[\hat H, \hat\rho\right] + \Gamma\sum_{\mathclap{\substack{\alpha=L,R\\n=g,e}}}q_{\alpha n}\mathcal{L}_{\alpha n}(\hat\rho)+\kappa\mathcal D[\hat a](\hat\rho).
\end{equation}
Here, the composite Hamiltonian \(\hat H\) reads $\hat H = \hat H_{\text{DQD}} + \hat H_{\text{cavity}} + g \sin\theta\left( \jump{e}{g}\hat a + \jump{g}{e}\hat a^\dagger \right)$ with the last term denoting the coherent light-matter interaction under the rotating-wave approximation of \(\hat H_{\text{DQD}-\text{cavity}}\). The second term on the right-hand-side of \Cref{eq:master_equation} describes the electron tunneling processes, with the corresponding transition weights \(q_{Le}=q_{Rg}=\cos^2(\theta/2),\ q_{Lg}=q_{Re}=\sin^2(\theta/2)\) and a strength parameter \(\Gamma\) which is taken to be constant under the wide-band approximation. \(\mathcal{L}_{\alpha n}\) is defined as
\begin{equation}
    \begin{split}
        \mathcal{L}_{\alpha n}(\hat\rho) :={}& f_\alpha(\epsilon_n)\mathcal{D}[\jump{n}{0}](\hat\rho) \\
        {}+{}&(1 - f_\alpha(\epsilon_n))\mathcal{D}[\jump{0}{n}](\hat\rho),
    \end{split}
\end{equation}
where \(f_\alpha(x) := 1/\left(1+e^{\beta_\alpha(x-\mu_\alpha)}\right)\) is the Fermi-Dirac distribution of lead \(\alpha\) characterized by an inverse temperature \(\beta_\alpha\) and a chemical potential \(\mu_\alpha\). The dissipator $\mathcal{D}[\hat A](\hat\rho)$ is given by
\begin{equation}
\mathcal{D}[\hat A](\hat\rho) := \hat A\hat \rho\hat A^\dagger - \frac{1}{2}\left\{\hat A^\dagger \hat A, \hat\rho\right\}.
\end{equation}
The third term on the right-hand-side of \Cref{eq:master_equation} accounts for photon loss from the cavity with the rate \(\kappa\). By including only the photon loss channel, the photonic reservoir is effectively treated as being at zero temperature~\cite{agarwalla_photon_2019}. For clarity, the details of the $9$ dissipation channels in \Cref{eq:master_equation}, along with the corresponding channel indices, current number and counting weights, are summarized in \Cref{tab:channel_numbering}.

To evaluate electronic and photonic currents in the present model, we define the following cumulative quantities
\begin{equation}\label{eq:N_12}
\mathcal N_1(t) := \sum_{j=1}^4 \nu_j n_j(t), \quad \mathcal N_2(t) := \nu_9 n_9(t).
\end{equation}
Here, \(\mathcal N_1(t)\) measures the net number of electrons transferred from the left lead to the DQD, obtained as a weighted sum over the stochastic counting processes \(n_j(t)\) which record jump events through the $j$th electronic channel up to time \(t\). Meanwhile, \(\mathcal N_2(t)\) counts the total number of jump events in the cavity photon loss channel. With \(\mathcal{N}_{1,2}\), the stationary electronic current $\mathcal{I}_1$ and the stationary photonic current $\mathcal{I}_2$ are then defined as $\mathcal I_1:=\left.\dd\mathbb E[\mathcal N_1]/\dd t\right|_{t\to\infty}$ and $\mathcal I_2:=\left.\dd\mathbb E[\mathcal N_2]/\dd t\right|_{t\to\infty}$, respectively, where $\mathbb E[\mathcal{O}]$ denotes the ensemble average of $\mathcal{O}$.

To analyze the counting statistics of these currents, we employ the full counting statistics formalism~\cite{levitov.93.JETP,Bagrets.03.PRB,Belzig.01.PRL,rammer_charge_representation_2004,esposito_fluctuation_2007,esposito_nonequilibrium_2009,landi_current_2024}.
This approach involves decomposing the density matrix into charge-specific density matrices (or \(\boldsymbol{\mathcal N}\)-resolved density matrices) \(\hat\varrho(\mathcal N_1,\mathcal N_2; t)\) according to the number of detected jumps \(\mathcal N_1,\mathcal N_2\)~\cite{Cook.81.PRA,Gurvitz.98.PRB,Flindt.04.PRB,Flindt.05.EPL}.
By performing a Fourier transform of the equations of motion of the charge-specific density matrices with respect to time-domain counting variables \(\mathcal N_{1,2}\), we can obtain the generalized quantum master equation for \(\hat\varrho(\chi_1,\chi_2; t)\)~\cite{Flindt.04.PRB,Flindt.05.EPL,esposito_nonequilibrium_2009,landi_current_2024}, where $\chi_{1,2}$ are the conjugate parameters (counting fields) corresponding to \(\mathcal N_{1,2}\) in the Fourier space.

For our system, the corresponding generalized quantum master equation can be obtained directly from \Cref{eq:master_equation} by replacing \(\hat\rho(t)\) with \(\hat\varrho(\chi_1,\chi_2; t)\) and modified each dissipator \(\mathcal{D}[\hat A](\hat\rho)\) to its tilted form
\begin{equation}
e^{i\chi_{\zeta}\nu_j}\hat A_j\hat \rho\hat A_j^\dagger - \frac{1}{2}\left\{\hat A_j^\dagger \hat A_j,\hat \rho\right\},
\end{equation}
where \(\chi_\zeta\) (with \(\zeta\) depending on \(j\), as specified in \Cref{tab:channel_numbering}) is the counting field associated with the \(\zeta\)-th current, and $\nu_j$ is the corresponding counting weight (see \Cref{tab:channel_numbering}). After this substitution, the superoperator acting on the counting field density operator \(\hat\varrho(\chi_1,\chi_2; t)\) is denoted as \(\mathcal L_{\boldsymbol{\chi}}\), where \(\boldsymbol{\chi}=(\chi_1,\chi_2)\), and is called the tilted Liouvillian superoperator. With the tilted Liouvillian superoperator,
the CGF of the stochastic variables \(\mathcal N_1\) and \(\mathcal N_2\) is defined as
\begin{equation}\label{eq:CGF}
\mathcal C(\chi_1,\chi_2; t) := \ln G(\chi_1,\chi_2; t),
\end{equation}
where the characteristic function \(G(\chi_1,\chi_2; t)\) is given by
\begin{equation}\label{eq:characteristic_function}
    G(\chi_1,\chi_2; t)=\tr\left[e^{t\mathcal L_{\boldsymbol{\chi}}}\hat\varrho(\chi_1,\chi_2; 0)\right].
\end{equation}
Here, ``$\tr$" denotes the trace operation.

The long-time behavior of fluctuations of the currents is captured by the scaled CGF (SCGF), defined as the time derivative of the CGF in the long-time limit,
\begin{equation}\label{eq:SCGF}
    \widetilde{\mathcal C}(\chi_1,\chi_2) := \lim_{t\to\infty} \frac{\partial}{\partial t} \mathcal C(\chi_1,\chi_2; t).
\end{equation}
THe SCGF $\widetilde{\mathcal C}(\chi_1,\chi_2)$ generates the scaled cumulants~\cite{landi_current_2024}, which correspond to the time derivative of cumulants in the long-time limit. The SCGF is closely related to the eigenvalues \(\{\lambda_1,\lambda_2,\dots\}\) of the tilted Liouvillian superoperator \(\mathcal L_{\boldsymbol{\chi}}\).
In most cases, if there exists a unique eigenvalue whose real part equals the spectral abscissa
\begin{equation}
    \Lambda:=\max\left(\{\re(\lambda_1),\re(\lambda_2),\dots\}\right),
\end{equation}
then the SCGF converges to this eigenvalue (see, e.g, Ref.~\cite{esposito_nonequilibrium_2009}).
This convergence is generally robust against variations in the initial state \(\hat\varrho(\chi_1,\chi_2; 0)\) or the block structure of the Jordan normal form of \(\mathcal L_{\boldsymbol{\chi}}\).
Using this property, we can efficiently evaluate the long-time limit of current cumulants by solving the eigenvalue problem of \(\mathcal L_{\boldsymbol{\chi}}\). For a detailed discussion of the sufficient and necessary conditions ensuring this convergence, we refer readers to \Cref{sec:convergence_of_scgf}.

\section{Relation between electron and photon statistics}\label{sec:2}
\begin{table}[b!]
\caption{Default parameter values. While varying the parameters, the default value will be denoted by subscript \(0\), e.g. \(\epsilon_0\).\label{tab:parameter_values}}
    \begin{tabularx}{\columnwidth}{l@{\extracolsep{\fill}}r}
    \toprule
    Light-matter coupling strength \(g\) & \(0.2050\si{\micro\electronvolt}\) \\
    Cavity frequency \(\omega\) & \(32.5\si{\micro\electronvolt}\) \\
    DQD elastic electron tunneling rate \(\tau\) & \(16.4\si{\micro\electronvolt}\) \\
    DQD Energy difference \(\epsilon\) & \(20\si{\micro\electronvolt}\) \\
    Electron reservoir tunneling rate \(\Gamma\) & \(16.56\si{\micro\electronvolt}\) \\
    Temperature \(1/\beta\) & \(8\si{\milli\kelvin}\) \\
    Cavity loss rate \(\kappa\) & \(0.0041\si{\micro\electronvolt}\) \\
    Chemical potential difference \(2\mu_c\) & \(100\si{\micro\electronvolt}\) \\
    \botrule
    \end{tabularx}
\end{table}
In this section, we analyze the joint statistics of electrons and photons utilizing the generalized quantum master equation derived from \Cref{eq:master_equation} as detailed in the previous section. We focus on scaled cumulants up to the second order which are well within the current experimental capabilities~\cite{Gabelli.09.PRB,Lebedev.16.PRB,Dasenbrook.16.PRL,Joshi.25.A}. These cumulants are defined as
\begin{equation}
    \begin{aligned}
        \mathcal I_j:={}&\left.-i\frac{\dd\widetilde{\mathcal C}}{\dd\chi_j}\right|_{\boldsymbol\chi=0} \quad &(j=1,2),\\
        \mathcal M_{n,m}:={}&\left.-\frac{\dd^2\widetilde{\mathcal C}}{\dd\chi_1^n\dd\chi_2^m}\right|_{\boldsymbol\chi=0}\quad&(m+n=2).
    \end{aligned}
\end{equation}
Here, $\mathcal{I}_1$ and $\mathcal{I}_2$ denote averaged electronic and photonic currents (first-order scaled cumulant), respectively. In the following, we will use the terms ``currents" and ``first-order cumulants" interchangeably. The set $\{\mathcal{M}_{n,m}\}$ represents the second-order scaled cumulants. Specifically, $\mathcal{M}_{2,0}$ and $\mathcal{M}_{0,2}$ quantify the electronic and photonic current noises, respectively, while $\mathcal{M}_{1,1}$ measures their cross-correlation. We adopt experimentally relevant parameter values from Refs.~\cite{liu_photon_2014, kulkarni_cavity-coupled_2014, agarwalla_photon_2019}, as detailed in \Cref{tab:parameter_values}. In subsequent parameter variations, the default values (denoted by a subscript $0$) are those provided in this table.

\begin{figure}[t!]
    \centering
    \includegraphics[width=\columnwidth]{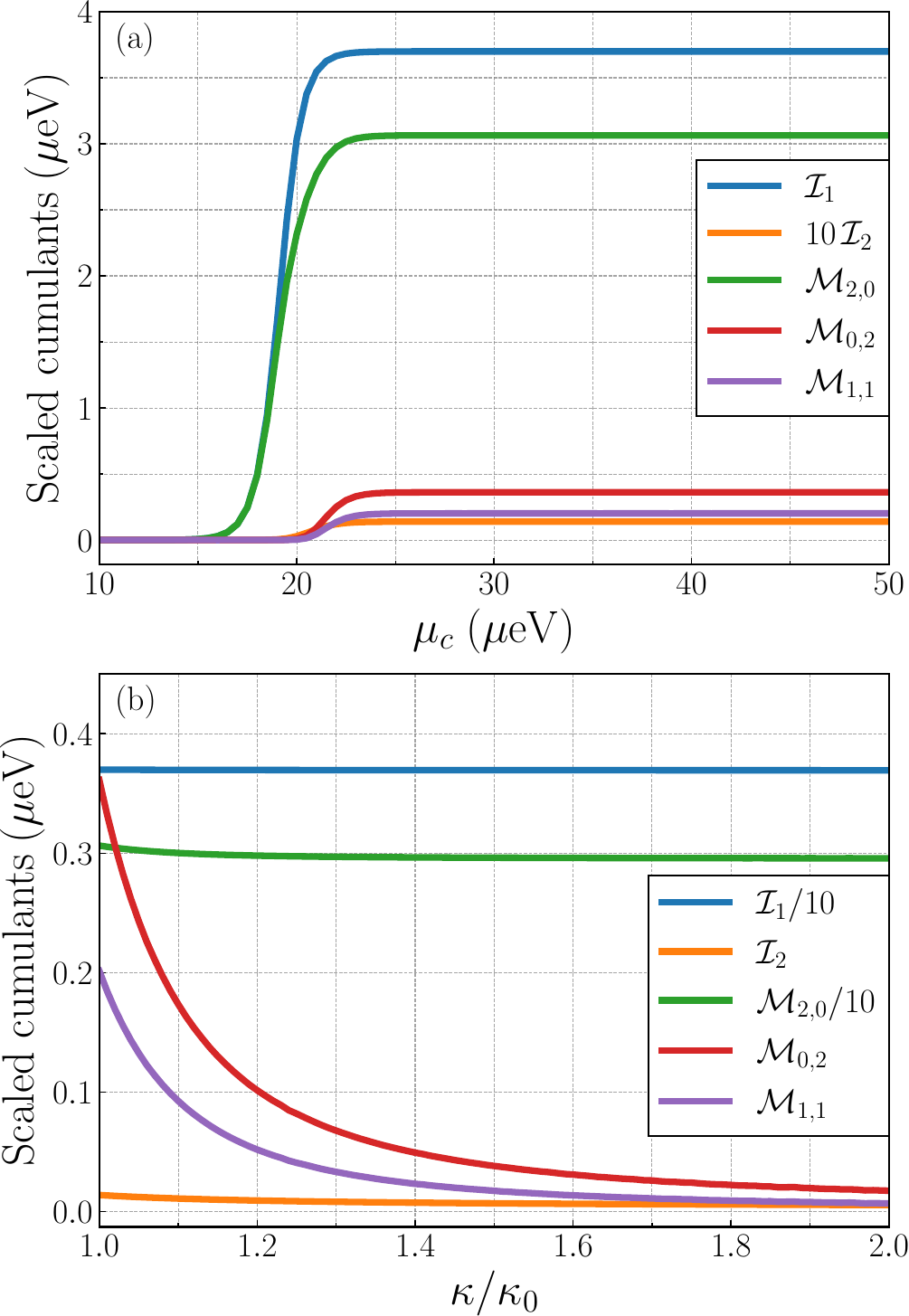}
    \caption{First-order and second-order scaled cumulants as functions of (a) chemical potential \(\mu_c\) and (b) photon loss rate \(\kappa\). In both plots, certain quantities are scaled by the factors indicated in the legends for a better illustration. All other parameters are set to their default values as specified in \Cref{tab:parameter_values}.\label{fig:mdep-kdep}}
    \refstepcounter{subfigure}\label{fig:mdep}\refstepcounter{subfigure}\label{fig:kdep}
\end{figure}

In the following, we symmetrically set the chemical potentials as \(\mu_c^L=-\mu_c^R=\mu_c>0\). This configuration leads to two distinct electron transport regimes: (i) When both \(\pm\Omega/2\) lie within the interval \(-\mu_c,+\mu_c\), the right lead absorbs electrons from both the ground and excited states (channels 5 and 7), while the left lead emits electrons towards both states (channels 2 and 4). (ii) When \(\pm\Omega/2\) lie outside \(-\mu_c,+\mu_c\), electrons in the ground state \(\ket{g}\) are mainly emitted by both leads (channels 2 and 6), and electrons in the excited state \(\ket{e}\) are primarily absorbed by both leads (channels 3 and 7); their inverse processes are largely suppressed due to energy barrier. The latter case thus corresponds to an electron transport process where both electronic and photonic currents are vanishingly small as can be seen from \Cref{fig:mdep} for small $\mu_c$. As $\mu_c$ increases, we observe that both the currents and the second-order scaled cumulants increase monotonically and eventually saturate at large $\mu_c$. In contrast, we find from \Cref{fig:kdep} that both the currents and the second-order scaled cumulants are monotonically decreasing functions of the photon loss rate \(\kappa\).
\begin{figure}[t!]
    \centering
    \includegraphics[width=\columnwidth]{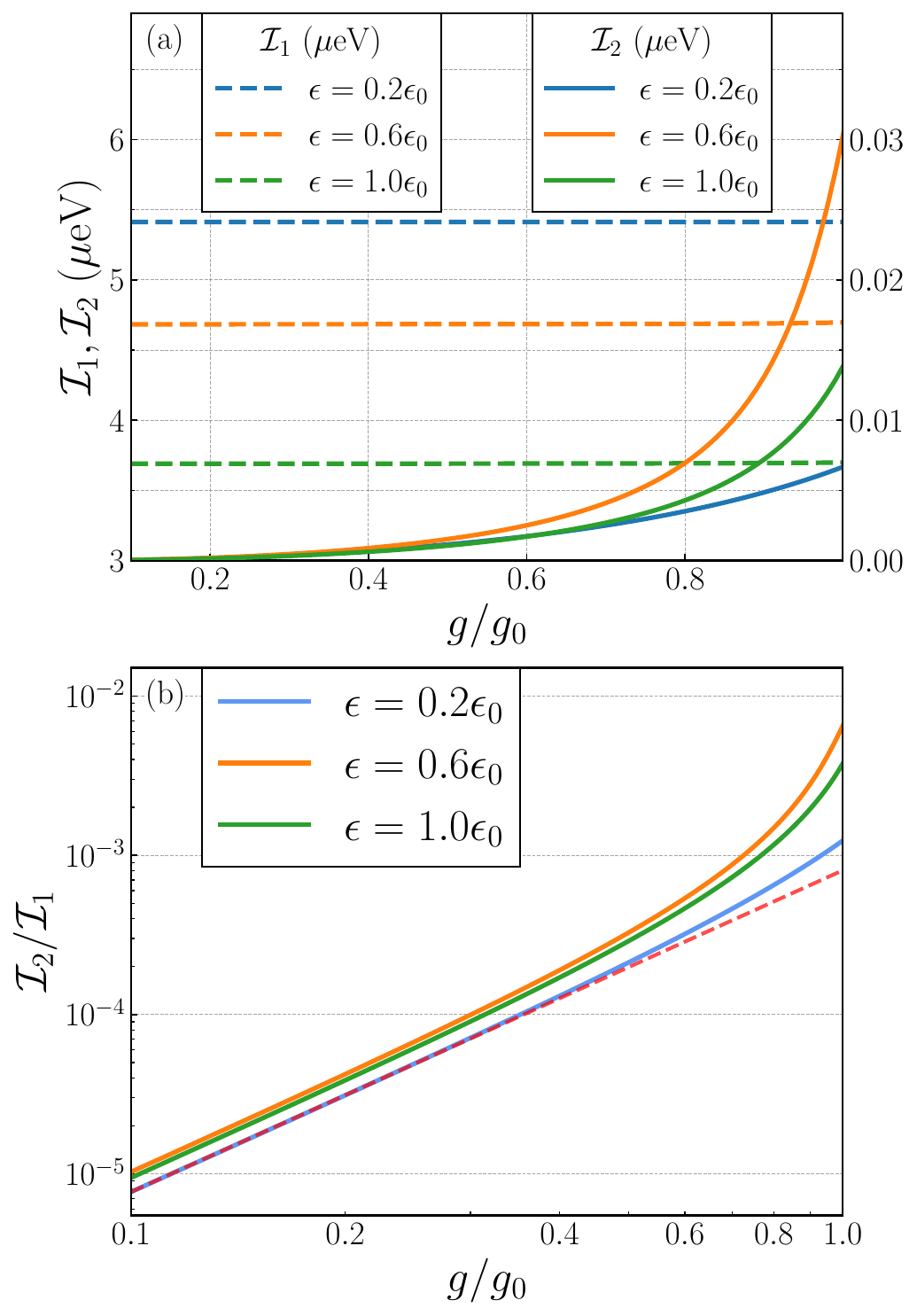}
    \caption{(a) Photonic ($\mathcal{I}_2$) and electronic ($\mathcal{I}_1$) currents and (b) their respective ratio $\mathcal{I}_2/\mathcal{I}_1$ as functions of the electron-photon coupling strength \(g\) for three values of energy splitting \(\epsilon\). The red dashed
    line in (b) serves as a guide to the eye for the power-law behavior $\mathcal{I}_2/\mathcal{I}_1\sim g^2$. All other parameters are set to their default values as specified in \Cref{tab:parameter_values}.\label{fig:ge-order1-with-ratio}}
    \refstepcounter{subfigure}\label{fig:ge-order1}\refstepcounter{subfigure}\label{fig:ge-order1-ratiolog}
\end{figure}

Now we examine the dependence of photonic and electronic currents on the electron-photon coupling strength \(g\). For this analysis, the chemical potential $\mu_c$ is set to its default value in \Cref{tab:parameter_values}, such that only channels 2 and 4 contribute to the electronic current flowing from the left lead. The electronic current $\mathcal I_1$ can then be expressed as
\begin{equation}\label{eq:I11}
    \begin{aligned}
        \mathcal I_1
        ={}&\sum_{\mathclap{j=2,4}}\nu_j\tr\left(\hat L_j^\dagger\hat L_j\hat\rho_{\text{ss}}\right)\\
        {}={}&\Gamma q_{Lg}\tr\left(\proj{0}\hat\rho_{\text{ss}}\right)+\Gamma q_{Le}\tr\left(\proj{0}\hat\rho_{\text{ss}}\right)\\
        {}={}&\Gamma \braket{0|\hat\rho_{\text{ss,e}}|0},
    \end{aligned}
\end{equation}
where \(\hat\rho_{\text{ss}}\) is the steady state at \(t\to\infty\) and \(\hat\rho_{\text{ss,e}}\) is the reduced electronic steady state.
This formula is consistent with the intuition that the current entering the DQD is determined by the probability for an empty DQD. The photonic current through channel 9 is given by
\begin{equation}\label{eq:I22}
    \mathcal I_2 = \nu_9\tr\left(\hat L_9^\dagger\hat L_9\hat\rho_{\text{ss}}\right) = \kappa\braket{\hat a^\dagger\hat a}_{\hat\rho_{\text{ss,p}}},
\end{equation}
where \(\hat\rho_{\text{ss,p}}\) is the reduced photonic steady state.

\begin{figure}[t!]
    \centering
    \includegraphics[width=\columnwidth]{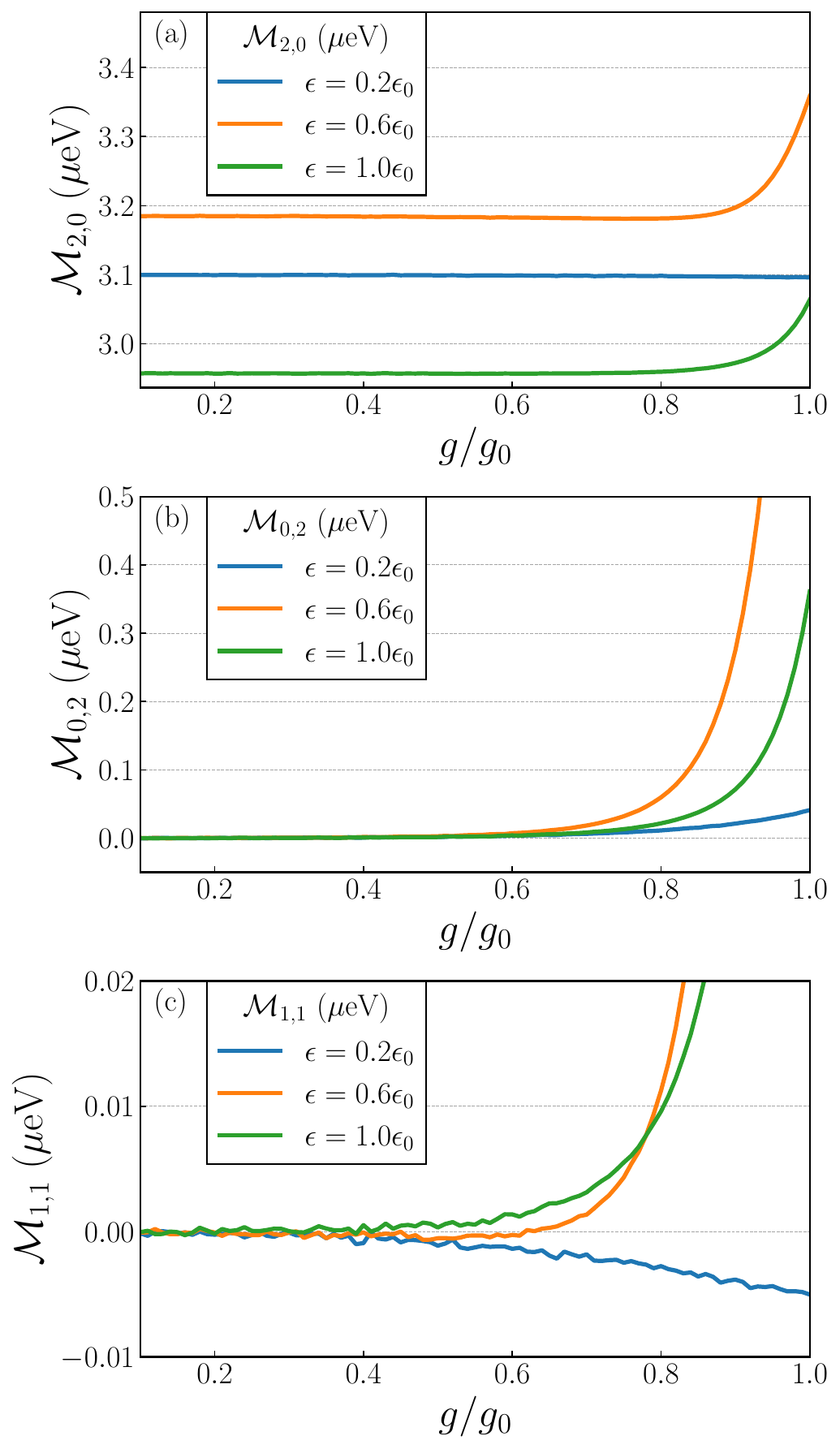}
    \caption{Second-order scaled cumulants versus coupling strength \(g\) for three values of energy splitting \(\epsilon\): (a) electronic current noise, (b) photonic current noise, and (c) electron-photon cross-correlation. All other parameters are set to their default values as specified in \Cref{tab:parameter_values}.\label{fig:ge-order2}}
    \refstepcounter{subfigure}\label{fig:ge20}\refstepcounter{subfigure}\label{fig:ge02}\refstepcounter{subfigure}\label{fig:ge11}
\end{figure}

We notice that the two currents $\mathcal{I}_{1,2}$ can be analytically connected. To proceed, we start from the Heisenberg equations of motion for the observables \(\proj{e}\) and \(\hat a^\dagger\hat a\) obtained using Eq. (\ref{eq:master_equation}),
\begin{align}
    \frac{\dd\proj{e}}{\dd t}
    ={}&{}-ig\sin\theta\left(\jump{e}{g}\hat a-\jump{g}{e}\hat a^\dagger\right)\nonumber\\
    {}&{}+\Gamma\left[\proj{0}\cos^2\left(\frac{\theta}{2}\right)-\proj{e}\sin^2\left(\frac{\theta}{2}\right)\right],\\
    \frac{\dd\hat a^\dagger\hat a}{\dd t}
    ={}&{}ig\sin\theta\left(\jump{e}{g}\hat a-\jump{g}{e}\hat a^\dagger\right)-\kappa\left(\hat a^\dagger\hat a\right).
\end{align}
Since time derivatives of the ensemble averages of \(\proj{e}\) and \(\hat a^\dagger\hat a\) vanish in the steady state limit, we readily find the following relation by noting the definitions given by Eqs. (\ref{eq:I11}) and (\ref{eq:I22}),
\begin{equation}\label{eq:excited-heisenberg}
    \mathcal{I}_2=P_0\cos^2\left(\frac{\theta}{2}\right)-P_e\sin^2\left(\frac{\theta}{2}\right).
\end{equation}
Here we use the notations \(P_0:=\braket{0|\hat\rho_{\text{ss,e}}|0}=I_1/\Gamma\), \(P_g:=\braket{g|\hat\rho_{\text{ss,e}}|g}\) and \(P_e:=\braket{e|\hat\rho_{\text{ss,e}}|e}\) for the steady state probabilities (diagonal elements) of the DQD system being empty, ground and excited, respectively.
Exploiting the Heisenberg equation of motion for $\proj{0}$ and taking the ensemble average in the steady state limit with \(\partial_t P_0=0\), we find
\begin{equation}\label{eq:empty-heisenberg}
    P_0=P_g\cos^2\left(\frac{\theta}{2}\right)+P_e\sin^2\left(\frac{\theta}{2}\right).
\end{equation}
Since $\sum_{i=0,g,e}P_i=1$, we can further eliminate \(P_g\) to obtain
\begin{equation}\label{eq:P_e-P_0}
    P_e=-\frac{1+3\sec\theta}{2}P_0+\frac{1+\sec\theta}{2}.
\end{equation}
Substituting \Cref{eq:P_e-P_0} into \Cref{eq:excited-heisenberg} yields a final relation between \(\mathcal{I}_1\) and \(\mathcal{I}_2\)
\begin{equation}\label{eq:current-relation-deterministic}
    \mathcal{I}_2=\mathcal{I}_1\frac{7+\cos2\theta}{8\cos\theta}-\frac{\Gamma}{4}\sin\theta\tan\theta.
\end{equation}
While \Cref{eq:I11,eq:I22,eq:current-relation-deterministic} are theoretical expressions for currents, their numerical results are obtained by resorting to the SCGF.

A set of representative results is depicted in \Cref{fig:ge-order1-with-ratio}. From \Cref{fig:ge-order1}, we observe that the averaged electronic current (dashed lines) increases only slightly with the electron-photon coupling strength \(g\).
This suggests that the probability for an empty DQD is largely insensitive to variations in \(g\), and we have also numerically verified that all diagonal elements of the electronic steady state are insensitive to \(g\) under this parameter setting.
In contrast, the photonic current (solid lines) rises markedly with the coupling strength \(g\), implying that a stronger coupling leads to a higher photon population in the cavity according to \Cref{eq:I22}. In \Cref{fig:ge-order1-ratiolog}, we further present results for the current ratio \(\mathcal I_2/\mathcal I_1\). We note that a prior study on a different electron-photon hybrid system reported that this current ratio exhibits a quadratic scaling with the coupling strength \(g\)~\cite{van_den_berg_charge-photon_2019}. Our results confirm that such a quadratic scaling (indicated by the red dashed line) holds in the weak-coupling regime, suggesting that this scaling behavior may be a general feature across various electron-photon hybrid systems. However, as shown in \Cref{fig:ge-order1-ratiolog}, the current ratio clearly deviates from the quadratic scaling at larger coupling strengths. This deviation becomes increasingly pronounced with larger energy splitting $\epsilon$. 

\begin{figure}[b!]
    \centering
    \includegraphics[width=\columnwidth]{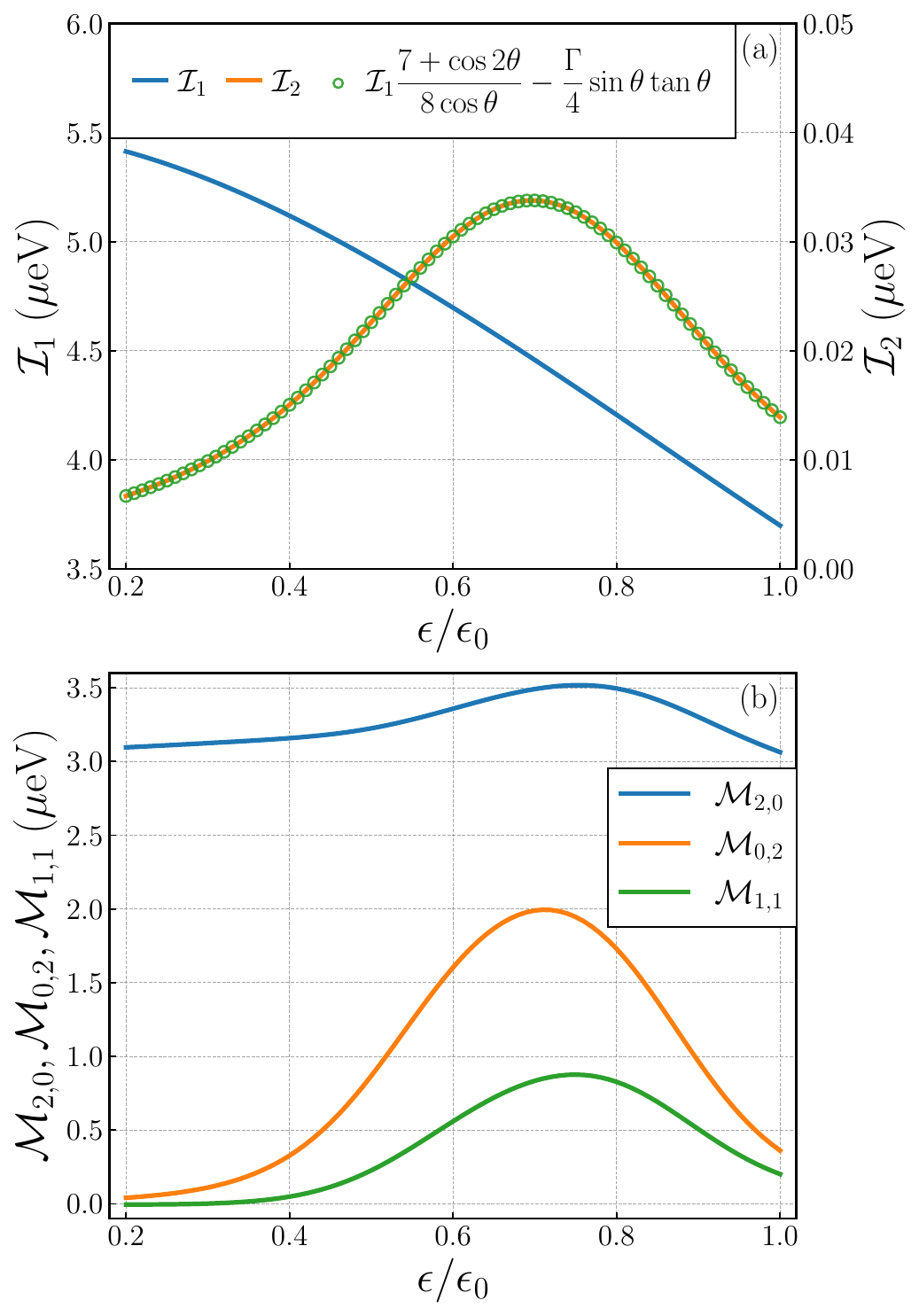}
    \caption{(a) \(\epsilon\)-dependence of first-order scale cumulants: Electronic current $\mathcal{I}_1$ (blue solid curve) and photonic current $\mathcal{I}_2$ (orange solid curve). We also verify the validity of Eq. (\ref{eq:current-relation-deterministic}) (green circles). (b) second-order scaled cumulants $\mathcal{M}_{2,0}$ (blue solid line), $\mathcal{M}_{0,2}$ (orange solid line) and $\mathcal{M}_{1,1}$ (green solid line) as functions of energy splitting \(\epsilon\). All other parameters are set to their default values as specified in \Cref{tab:parameter_values}.\label{fig:edep}}
    \refstepcounter{subfigure}\label{fig:edep-order1}\refstepcounter{subfigure}\label{fig:edep-order2}
\end{figure}

In \Cref{fig:ge-order2}, we turn to the $g$-dependence of the second-order scaled cumulants $\mathcal{M}_{2,0}$, $\mathcal{M}_{0,2}$ and $\mathcal{M}_{1,1}$ that represent the electronic current noise, photonic current noise and electron-photon cross correlation, respectively. A comparison of the results in \Cref{fig:ge20} and \Cref{fig:ge02} reveals that the photonic current noise is an order of magnitude smaller than the electronic current noise. We attribute this difference in current noise to the fact that only a single channel contributes to both the photonic current and its second-order fluctuations (see \Cref{tab:channel_numbering}). The electron-photon cross-correlation, presented in \Cref{fig:ge11}, becomes negative at large coupling strengths when the energy splitting $\epsilon$ is relatively small, indicating anti-correlated electronic and photonic currents. As the energy splitting increases, the cross-correlation turns positive across the entire range of coupling strengths. This transition can be understood by considering the behavior of the (conditional) probability \(\sin^2(\theta/2)\), where $\theta = \arctan(2\tau/\epsilon)$.
This probability--which corresponds to the right lead absorbing an excited electron (channel 7) and the left lead emitting a ground-state electron (channel 2)--decreases from approximately \(44\%\) (for $\epsilon=0.2\epsilon_0$) to about \(24\%\) (for $\epsilon=1.0\epsilon_0$). The former channel suppresses photon emission by depleting excited electrons, while the latter suppresses it by populating the electronic ground state of the DQD system. We note that Ref.~\cite{van_den_berg_charge-photon_2019} also reported a quadratic scaling of the correlation ratio \(\mathcal M_{1,1}/\mathcal M_{2,0} \propto g^2\). However, as shown in in \Cref{fig:ge-order2}, our model deviates significantly from this scaling, particularly since the ratio \(\mathcal M_{1,1}/\mathcal M_{2,0}\) can become negative for small energy splittings, in contrast to the current ratio, which remains positive and follows quadratic scaling in the weak-coupling regime.

\begin{figure*}[t!]
    \centering
    \includegraphics[width=\textwidth]{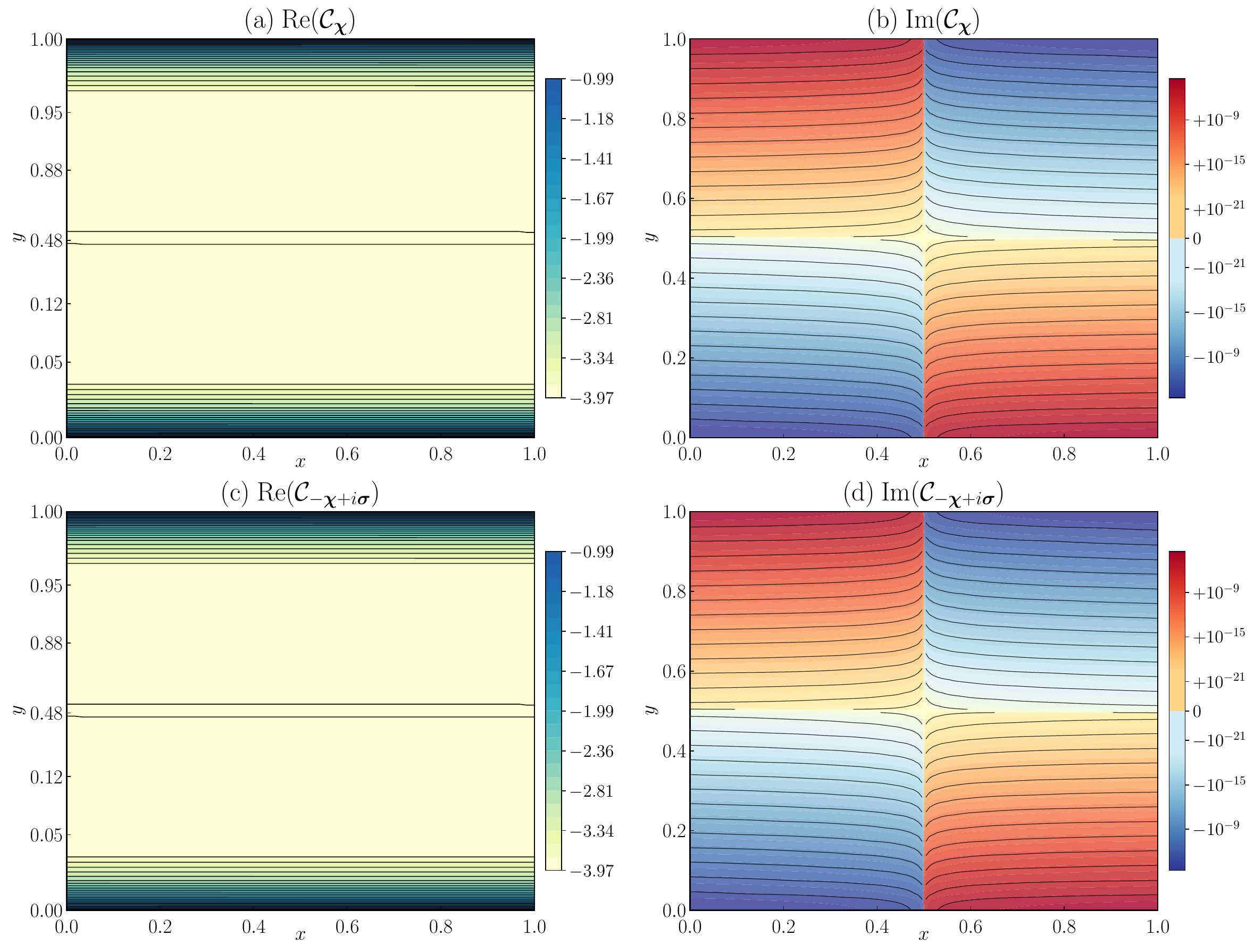}
    \caption{Real and imaginary part of the scaled cumulant generating functions on a two-dimensional plane passing through the fluctuation symmetric point $i\boldsymbol\sigma/2$ of $\boldsymbol\chi$ and $-\boldsymbol\chi+i\boldsymbol\sigma$ appeared in \Cref{eq:symmetry_scgf}, using a modified quantum Lindblad master equation \Cref{eq:modified_master}. Left panel: Real parts of (a) $\mathcal{C}_{\boldsymbol\chi}$ and (c) $\mathcal C_{-\boldsymbol\chi+i\boldsymbol\sigma}$. Right panel: Imaginary parts of (b) $\mathcal{C}_{\boldsymbol\chi}$ and (d) $\mathcal C_{-\boldsymbol\chi+i\boldsymbol\sigma}$. $x$ and $y$ are introduced as parameters such that the first four counting fields are parameterized as \((0.1+i\boldsymbol\sigma_{1-4}/2)x\), while the last counting field is given by \((0.1+i\sigma_{5}/2)y\). We set the chemical potential of the photonic reservoir \(\mu_p=32.02\,\mu\text{eV}\), other parameters are set to their default values as specified in \Cref{tab:parameter_values}.\label{fig:SCGF-plane}}
    \refstepcounter{subfigure}\label{fig:re-dim3-prec40}\refstepcounter{subfigure}\label{fig:im-dim3-prec40}
\end{figure*}
Previous results indicate that both the first- and second-order scaled cumulants exhibit sensitivity to the energy splitting \(\epsilon\). To further elucidate this behavior, we show in \Cref{fig:edep} the first- and second-order scaled cumulants as functions of the energy splitting \(\epsilon\). From \Cref{fig:edep-order1}, we observe that the electronic current decreases monotonically with $\epsilon$, whereas the photonic current displays a non-monotonic dependence, albeit with a much smaller magnitude. We note that the peak in the photonic current does not stem from electron-photon resonances, as the DQD energy splitting satisfies \(\Omega>\omega\) across the entire range of $\epsilon$ considered. According to Eqs. (\ref{eq:I11}) and (\ref{eq:I22}), the monotonic decrease in the electronic current implies that the electronic population of the empty state in the DQD system decreases with $\epsilon$, whereas the cavity photon occupation varies non-monotonically. Given the relation between $\mathcal{I}_1$ and $\mathcal{I}_2$ in \Cref{eq:current-relation-deterministic}, it suffices to understand the monotonic decrease of $P_0$ with increasing $\epsilon$. Based on Eq. (\ref{eq:P_e-P_0}), we 
treat $P_0$ as an implicit function \(P_0(P_e(\epsilon),\theta(\epsilon))\) and analyze its variation with $\epsilon$ through $P_e(\epsilon)$ and $\theta(\epsilon)$. We first note that $P_0$ is a decreasing function of $P_e$ as
\begin{equation}
    \frac{\partial}{\partial P_e}P_0(P_e,\theta) ~=~ -\frac{2\cos\theta}{3+\cos\theta} \;<\; 0.
\end{equation}
In the parameter regime considered, $P_e$ increases with $\epsilon$: A larger \(\epsilon\) enhances channels 4 and 7 while suppressing channels 2 and 5, due to the increase in \(\cos^2(\theta/2)\) and decrease in \(\sin^2(\theta/2)\), reflecting a stronger preference for excited electrons over ground-state electrons. 
This increasing trend of \(P_e(\epsilon)\) is confirmed numerically. Therefore, the $P_e$-dependence of $P_0$ tends to reduce $P_0$ as $\epsilon$ increases. We next examine the dependence of $P_0$ on $\theta$ and behaviors of $\theta(\epsilon)$. From \Cref{fig:edep-order1}, we observe that \(\mathcal{I}_1<\Gamma/3=5.52\si{\micro\electronvolt}\), implying \(P_0<1/3\), or equivalently \(P_e>1/3\). Thus,
\begin{equation}
    \frac{\partial}{\partial \theta}P_0(P_e,\theta) ~=~ \frac{2\sin\theta}{(3+\cos\theta)^2}(3 P_e-1) > 0.
\end{equation}
Namely, $P_0$ is a increasing function of \(\theta\). However, $\theta$ decreases monotonically with $\epsilon$ over the range considered in \Cref{fig:edep} since $\frac{\partial\theta}{\partial \epsilon} =-2\tau/(\epsilon+2\tau)^2 < 0$. Hence, the variation on $\theta$ with $\epsilon$ also acts to reduce $P_0$. Together, the dependencies on $P_e(\epsilon)$ and $\theta(\epsilon)$ yield a net monotonic decrease in \(P_0\) as $\epsilon$ increases, resulting in the decreasing trend of \(\mathcal{I}_1\) shown in \Cref{fig:edep-order1}.

\section{Recovering fluctuation symmetry}\label{sec:3}
In the previous section, we analyzed the joint statistics of electrons and photons up to the second order using a conventional quantum master equation [cf. \Cref{eq:master_equation}]. However, a fundamental question that is still not settled is whether this conventional quantum master equation [cf. \Cref{eq:master_equation}] satisfies the fluctuation symmetry enforced by the fluctuation theorem~\cite{esposito_fluctuation_2007,esposito_nonequilibrium_2009,Jarzynski.11.AR,Seifert.12.RPP,Andrieux.09.NJP,Saito.08.PRB}, thus is consistent with nonequilibrium thermodynamics. At the level of CGF, this fluctuation symmetry dictates the following relation~\cite{esposito_fluctuation_2007,esposito_nonequilibrium_2009,Jarzynski.11.AR,Seifert.12.RPP,Andrieux.09.NJP,Saito.08.PRB}
\begin{equation}\label{eq:symmetry_scgf}
	\mathcal C_{\boldsymbol\chi}=\mathcal C_{-\boldsymbol\chi+i\boldsymbol\sigma},
\end{equation}
The components of the vector \(\boldsymbol\sigma\) are commonly referred to as thermodynamic affinities, which are determined by reservoir parameters. From this fluctuation symmetry, well-known results such as the Onsager reciprocity relations or fluctuation-dissipation relations can be recovered~\cite{Saito.08.PRB,Andrieux.09.NJP}. It is therefore important to examine whether this fluctuation symmetry is preserved within the framework of quantum master equations which involve approximations to some extent~\cite{Segal.18.PRE,Liu.21.PRE,Gerry.23.PRE}.

We aim for a general understanding of the sufficient and necessary conditions under which this fluctuation symmetry holds within the framework of the quantum Lindblad master equations tailored for hybrid quantum systems. Our analysis follows and extends the approach of Ref.~\cite{landi_current_2024}, with detailed derivations provided in \Cref{sec:condition_for_fluctuation_theorem} for clarity. In our analysis, we assume that all jump operators \(\hat L_j\) corresponding to channel \(j\) and the Hamiltonian \(\hat H\) are real matrices in some basis. This assumption, which is obviously valid in our model, reflects the presence of certain time-reversal symmetry~\cite{sakurai_symmetry_2020,landi_current_2024}.
We demonstrate that the fluctuation symmetry [cf. \Cref{eq:symmetry_scgf}] is satisfied by quantum Lindblad master equation if and only if the following two conditions are simultaneously met. The first condition [cf. \Cref{eq:zero_weight_condition_even}] reads
\begin{equation}\label{eq:zero_weight_condition_even_simp}
    \sum_j L_j^T\otimes L_j^T=\sum_j L_j\otimes L_j,
\end{equation}
where both two summations are performed over all channels that are not counted (i.e., those with no associated counting fields or zero counting weights) and \(T\) denotes the matrix transpose.

The second condition [cf. \Cref{eq:other_weight_condition_even}] applies to counted channels. For a given counting field \(\chi_\iota\) (where \(\iota\) takes value from the current numbers in \Cref{tab:channel_numbering} or \Cref{tab:new_channel_numbering}), let \(Z_\iota:=\{\nu_\eta\;|\;\eta=1,2,\dots,F\}=\{\nu_j\;|\;\nu_j>0\;\land\;j=1,2,\dots\;\land\;\zeta_j=\iota\}\) denote the set of \(F\) distinct positive counting weights associated with the counting field \(\chi_\iota\). We require that all counting weights form a symmetric set around zero, that is, for every positive counting weight in \(Z_\iota\), there exists a corresponding negative weight in a complementary set \(-Z_\iota\) and vice versa. The second condition [cf. \Cref{eq:other_weight_condition_even}] can then be stated as follows: For each pair \(\pm \nu_\eta\), the following positive proportionality must hold
\begin{equation}\label{eq:other_weight_condition_even_simp}
	\sum_{j}L_j^T\otimes L_j^T
    \;=\;
    p_\eta\sum_{j'}L_{j'}\otimes L_{j'}.
\end{equation}
Here, $p_\eta>0$, the summation on the left (right) includes all channels associated with \(\chi_\iota\) that carry the counting weight \(\nu_\eta\) (\(-\nu_\eta\)).
We also require that for a fixed \(\iota\), the values of \(\ln p_\eta/\nu_\eta\) are independent of \(\eta\), and this common value defines the \(\iota\)-th affinity \(\sigma_\iota=\ln p_\eta/\nu_\eta\) that enters the fluctuation symmetry in \Cref{eq:symmetry_scgf}. In the special case where only a single channel is involved, \Cref{eq:other_weight_condition_even_simp} reduces to a detailed balance condition~\cite{landi_current_2024} (see also \Cref{eq:detailed_balance} in \Cref{sec:condition_for_fluctuation_theorem}).

\begin{table*}[t!]
\caption{Jump channel definitions for the modified quantum master equation \Cref{eq:modified_master} which includes a photon excitation process.\label{tab:new_channel_numbering}}
	\begin{tabular}{|c|c|c|c|c|c|c|c|c|c|c|c|}
		\hline
		Channel Number \(j\) & 1 & 2 & 3 & 4 & 5 & 6 & 7 & 8 & 9 & 10 \\
		\hline
		Related reservoir & \multicolumn{4}{c|}{DQD left lead} & \multicolumn{4}{c|}{DQD right lead} & \multicolumn{2}{c|}{Cavity} \\
		\hline
		Jump Operator (coefficient omitted) & \(\jump{0}{g}\) & \(\jump{g}{0}\) & \(\jump{0}{e}\) & \(\jump{e}{0}\)  & \(\jump{0}{g}\) & \(\jump{g}{0}\) & \(\jump{0}{e}\) & \(\jump{e}{0}\) & \(\ \hat a\ \) & \(\hat a^\dagger\) \\
		\hline
        Coefficient & \multicolumn{8}{c|}{Same as \Cref{eq:master_equation}}  & \(\kappa (1+n_{\rm{th}})\) & \(\kappa n_{\rm{th}}\) \\
        \hline
		Current number \(\zeta_j\) & \multicolumn{2}{c|}{\(1\)} & \multicolumn{2}{c|}{\(2\)} & \multicolumn{2}{c|}{\(3\)} & \multicolumn{2}{c|}{\(4\)} & \multicolumn{2}{c|}{\(5\)} \\
		\hline
		Counting weight \(\nu_j\) & \(-1\) & \(1\) & \(-1\) & \(1\) & \(-1\) & \(1\) & \(-1\) & \(1\) & \(1\) & \(-1\) \\
		\hline
	\end{tabular}
\end{table*}

We find that the conventional quantum master equation \Cref{eq:master_equation} fails to satisfy both conditions specified in \Cref{eq:zero_weight_condition_even_simp,eq:other_weight_condition_even_simp}. This failure stems from two main reasons: (i) Only a subset of channels is included in the previous counting statistics (as summarized in \Cref{tab:channel_numbering}), which prevents the full set of affinities required by the fluctuation symmetry from being defined, and (ii) the photonic component of \Cref{eq:master_equation} incorporates only a loss channel, whereas the electronic part accounts for both excitation and decay processes. To satisfy the condition \eqref{eq:other_weight_condition_even_simp}, the conventional quantum master equation must be modified by introducing a photon excitation process that balances the photon loss. This modification corresponds to coupling the cavity to a thermal photon reservoir at finite temperature, thereby enforcing detailed balance for both electronic and photonic subsystems in the hybrid electron-photon system. The modified quantum master equation takes the following form
\begin{equation}\label{eq:modified_master}
    \begin{split}
        \frac{d\hat\rho}{dt} = -{}&i\left[\hat H, \hat\rho\right] + \Gamma\,\sum_{\mathclap{\substack{\alpha=L,R\\n=g,e}}}\,q_{\alpha n}\mathcal{L}_{\alpha n}(\hat\rho)\\
        +{}& \kappa(1+n_{\rm{th}})\mathcal{D}[\hat a](\hat\rho) + \kappa n_{\rm{th}}\mathcal{D}[\hat a^\dagger](\hat\rho),
    \end{split}
\end{equation}
where \(n_{\rm{th}}:=[e^{\beta(\omega-\mu_p)}-1]^{-1}\) denotes the averaged thermal photon occupation number in the cavity, and we assume that the photonic reservoir shares the same temperature as  the electronic reservoirs. 

\begin{figure}[b!]
    \centering
    \includegraphics[width=\columnwidth]{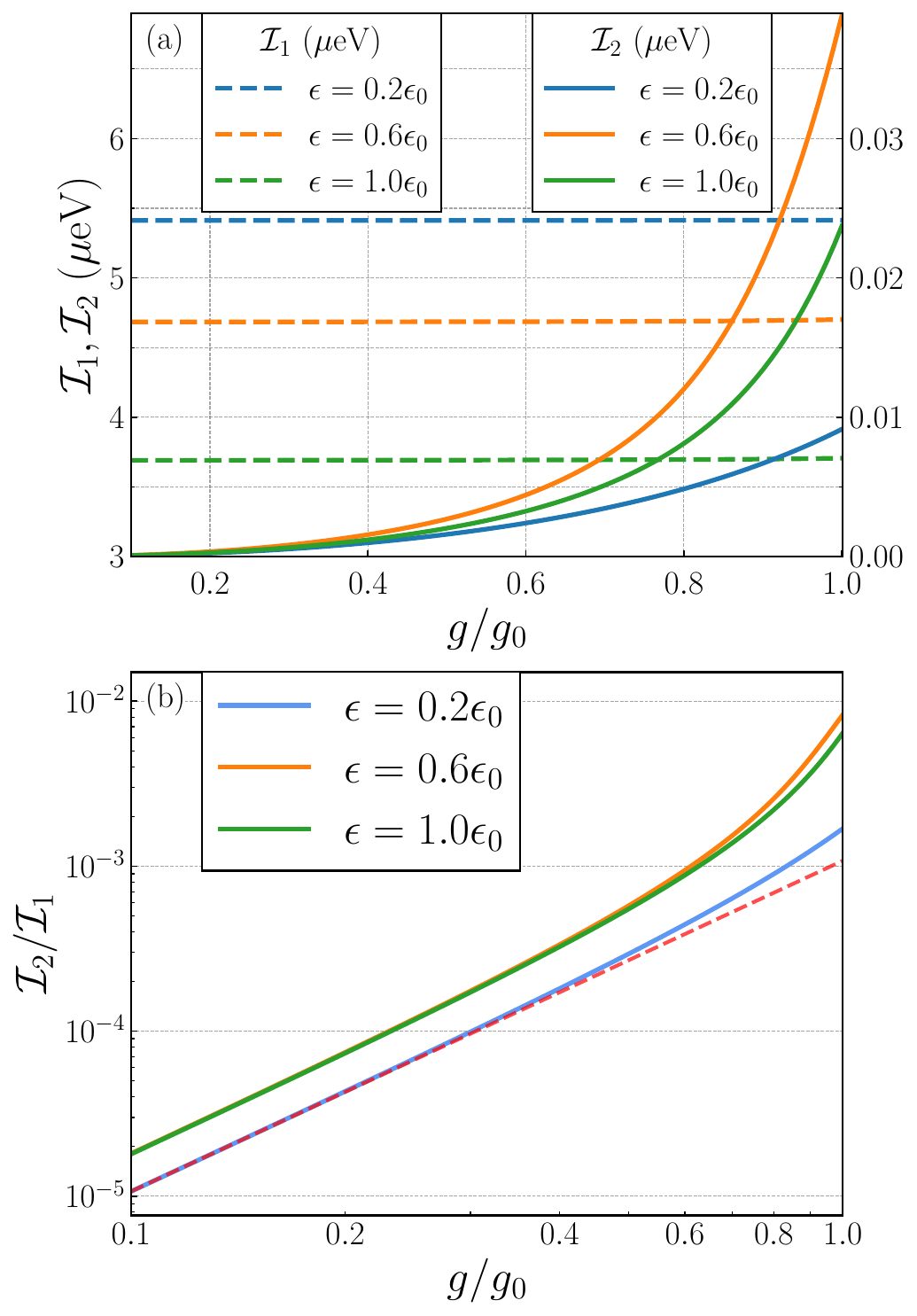}
    \caption{(a) Photonic ($\mathcal{I}_2$) and electronic ($\mathcal{I}_1$) currents and (b) their respective ratio $\mathcal{I}_2/\mathcal{I}_1$ as functions of the electron-photon coupling strength \(g\) for three values of energy splitting \(\epsilon\). We obtain the numerical results by using the modified quantum Lindblad master equation \Cref{eq:modified_master}. The red dashed
    line in (b) serves as a guide to the eye for the power-law behavior $\mathcal{I}_2/\mathcal{I}_1\sim g^2$. \(\mu_p=32.02\,\mu\text{eV}\), other parameters are set to their default values as specified in \Cref{tab:parameter_values}.\label{fig:ge-order1-balanced}}
\end{figure}

To satisfy the condition \eqref{eq:zero_weight_condition_even_simp}, we group all channels into five pairs, with each pair associated with a counting field (see details summarized in \Cref{tab:new_channel_numbering}). Consequently, we introduce five independent counting fields to the modified quantum master equation \Cref{eq:modified_master}, ensuring that every channel is counted--in direct contrast to the previous modeling which employed only two counting fields. Among these five counting fields, each electronic lead is associated with two: one for the jump process $|0\rangle\leftrightarrow|g\rangle$ and the other for $|0\rangle\leftrightarrow|e\rangle$. The remaining counting field corresponds to photon exchange. We remark that, with the inclusion of a photon excitation process, each channel pair can now be assigned a symmetric counting weight set \(\{+1,-1\}\), regardless of the type of particles. By doing so, the condition \eqref{eq:other_weight_condition_even_simp} can be satisfied with only one value \(Z_\iota=\{1\}\) for each counting field \(\chi_\iota\) and thus the logarithm of these positive coefficients \(\ln p_1=\sigma_\iota\) will be the affinities in the fluctuation symmetry \eqref{eq:symmetry_scgf}. We denote $\mu_p$ as the chemical potential of the photon reservoir, then the corresponding five affinities read $\sigma_1 = -\beta(-\Omega/2-\mu_c)$, $\sigma_2 = -\beta(+\Omega/2-\mu_c)$, $\sigma_3 = -\beta(-\Omega/2+\mu_c)$, $\sigma_4 = -\beta(+\Omega/2+\mu_c)$ and $\sigma_5 = -\beta(+\omega_c-\mu_p)$.

To demonstrate that the modified quantum Lindblad master equation \Cref{eq:modified_master} satisfies the fluctuation symmetry, we compare both the real and imaginary parts of the SCGF $\mathcal{C}_{\boldsymbol\chi}$ and $\mathcal C_{-\boldsymbol\chi+i\boldsymbol\sigma}$ in \Cref{fig:SCGF-plane}. The two-dimensional plane in this figure is parameterized by two real variables \(x\) and \(y\) for a better illustration: The first four counting fields are parameterized as \((0.1+i\boldsymbol\sigma_{1-4}/2)x\), while the last counting field is given by \((0.1+i\sigma_{5}/2)y\). As can be seen from the figure, we observe that both the real and imaginary parts of $\mathcal{C}_{\boldsymbol\chi}$ and $\mathcal C_{-\boldsymbol\chi+i\boldsymbol\sigma}$ exhibit excellent agreement, thereby confirming that the fluctuation symmetry is indeed satisfied by our modified quantum Lindblad master equation \Cref{eq:modified_master}.

Now that the new SCGF satisfies the fluctuation symmetry, we revisit the dependence of the electronic current $\mathcal{I}_1$ and photonic current $\mathcal{I}_2$ on the electron-photon coupling strength $g$, and examine whether the quadratic scaling of $\mathcal{I}_2/\mathcal{I}_1$ with respect to $g$ in the weak coupling regime persists under this symmetry. In \Cref{fig:ge-order1-balanced} (a) and (b), we present numerical results for currents $\mathcal{I}_{1,2}$ and their ratio $\mathcal{I}_2/\mathcal{I}_1$ as  computed from the new SCGF, respectively. Compared with the results showed in \Cref{fig:ge-order1-with-ratio}, we observe that the currents still depict similar trends as the coupling strength $g$ varies. Particularly, we find that the ratio $\mathcal{I}_2/\mathcal{I}_1$ continues to scale quadratically with $g$ in the weak coupling regime. We have also checked that the $g$-dependence of the second-order scaled cumulants shows no qualitative difference from the previous model. Our results thus indicate that although the conventional master equation description in Eq.~\eqref{eq:master_equation} fails to satisfy the fluctuation symmetry, it still provides a good approximation at the relatively low temperature considered (see Table~{\ref{tab:parameter_values}}), yielding qualitatively similar cumulant behaviors.


\section{Conclusion}\label{sec:4}
We have analyzed the full counting statistics of an experimentally feasible electron-photon hybrid system composed of a double quantum dot coupled to an optical cavity. By incorporating both electron and photon counting fields into the quantum Lindblad master equation, we investigated the joint statistical behavior of electronic and photonic currents. Our numerical results revealed distinct characteristics for these currents: while the electronic current exhibits weak dependence on the coupling strength $g$, the photonic current increases monotonically with $g$. Notably, we observed anti-correlated electron and photon currents at small energy splitting, resulting from the competition between electronic transport and photonic emission processes. Furthermore, we confirmed that the ratio of photonic to electronic currents follows a quadratic scaling with 
$g$ in the weak coupling regime, consistent with earlier theoretical predictions~\cite{van_den_berg_charge-photon_2019}. In contrast, the ratio between their second-order fluctuations does not adhere to the same quadratic scaling, diverging from the behavior anticipated in Ref.~\cite{van_den_berg_charge-photon_2019}.

We have further established the conditions under which the joint statistics satisfy the fluctuation symmetry~\cite{esposito_fluctuation_2007,esposito_nonequilibrium_2009,Jarzynski.11.AR,Seifert.12.RPP,Andrieux.09.NJP,Saito.08.PRB}. Based on these conditions, we noted that a conventional quantum Lindblad master equation for such hybrid systems~\cite{agarwalla_photon_2019} fails to satisfy the fluctuation symmetry, as it omits the photon gain channel by treating the photonic reservoir as being at zero temperature. To restore the fluctuation symmetry in the joint statistics, we modified the quantum Lindblad master equation to include both photon loss and gain channels, effectively representing the photonic reservoir at finite temperature. We numerically verified that the cumulant generating function derived from the modified master equation satisfies the fluctuation symmetry. Our results thus offer fundamental insights into fluctuation behavior in electron-photon hybrid quantum systems and provide practical guidance for constructing quantum master equations that are consistent with nonequilibrium quantum thermodynamics.



\section*{Acknowledgments}
J.L. acknowledges support from the National Natural Science Foundation of China (Grant No. 12205179), the Shanghai Pujiang Program (Grant No. 22PJ1403900) and start-up funding of Shanghai University.

\appendix
\crefalias{section}{appendix}

\renewcommand{\theequation}{A\arabic{equation}}
\setcounter{equation}{0}  
\section{Convergence of SCGF}\label{sec:convergence_of_scgf}
In this appendix, we analyze the necessary and sufficient conditions for the convergence of the SCGF given by \Cref{eq:SCGF} in the main text for quantum state spaces with a finite dimension \(d\) in detail.

For a finite-dimensional system, we take the vectorized form of \(\mathcal L_{\boldsymbol\chi}\) and the initial counting field density operator \(\hat\varrho(\boldsymbol\chi; 0)\).
Here, \(\boldsymbol\chi\) is a list of counting fields whose length is not limited to the case of two counting fields.
By choosing a basis where \(\mathcal L_{\boldsymbol\chi}\) is in its Jordan canonical form, we can write
\begin{equation}
    \mathcal L_{\boldsymbol\chi}=
    \bigoplus_{j=1}^\Xi\left(\lambda_j\mathbb I_j+N_j\right),\quad e^{t\mathcal L_{\boldsymbol\chi}}
    =
    \bigoplus_{j=1}^\Xi
    e^{t\lambda_j}e^{tN_j}.
\end{equation}
Here, \(\{\lambda_j\}\) are the eigenvalues of $\mathcal L_{\boldsymbol\chi}$, \(\mathbb I_j\) is an \(r_j \times r_j\) identity matrix and \(\oplus\) represents matrix direct sum.
\(N_j\) is an upper shift matrix, i.e. the superdiagonal elements are all ones and the rest are zeros.
The second equality directly follows from the series definition of the matrix exponential~\cite{hall_lie_2015}.
Consequently, the characteristic function \Cref{eq:characteristic_function} in the main text can be expressed as
\begin{equation}\label{eq:G_expression}
    \begin{split}
    G(t)
    =&
    \bbracckett{\tr}{\bigoplus_{j=1}^\Xi e^{t\lambda_j}e^{tN_j}}{\varrho}\\
    =&
    \sum_{j=1}^\Xi e^{t\lambda_j}\bbracckett{\tr^{(j)}}{e^{tN_j}}{\varrho^{(j)}}.
    \end{split}
\end{equation}
Here \(\bbra{\tr}\) is a vectorized matrix trace transformed by the basis change made previously for a Jordan normal form of \(\mathcal L_{\boldsymbol\chi}\).
The vectorized matrix trace before the basis change has ones at the positions \(1+(k-1)(d+1),k=1,2,\dots,d\) and zeros elsewhere, i.e. stacking columns of an identity matrix.
\(\varrho^{(j)}\) is a vector in the subspace corresponding to the \(j\)-th Jordan block of \(\mathcal L_{\boldsymbol\chi}\).
That is, if the \(j\)-th Jordan block is indexed from \(s:=1+\sum_{r=1}^{j-1}n_r\) to \(s+r_j-1\), then \(\varrho^{(j)}_k=\varrho_{s+k-1}\) where \(k=1,2,\dots,r_j\).
The sliced trace \(\bbra{\tr^{(j)}}\) is defined similarly to \(\varrho^{(j)}\).

The inner product \(\bbracckett{\tr^{(j)}}{e^{tN_j}}{\varrho^{(j)}}\) is a polynomial in \(t\)
\begin{equation}
    \begin{split}
        \mathcal W_j
        :=&
        \bbracckett{\tr^{(j)}}{e^{tN_j}}{\varrho^{(j)}}\\
        =&
        \bbracckett{\tr^{(j)}}{\sum_{k=0}^{r_j-1}\frac{t^k}{k!}N_j^k}{\varrho^{(j)}}\\
        =&
        \sum_{k=0}^{r_j-1}\frac{t^k}{k!}\sum_{l=1}^{r_j-k}\tr^{(j)}_{l}\varrho^{(j)}_{l+k}.
    \end{split}
\end{equation}
Note that the coefficient \(\sum_{l=1}^{r_j-k}\tr^{(j)}_{l}\varrho^{(j)}_{l+k}\) is simply the sum of \(k\)-th superdiagonal elements of the outer product \(\jump{\tr^{(j)}}{\varrho^{(j)}}\), so the order \(Q_j\) of the polynomial is the index of the nonzero \(k\)-th superdiagonal with largest \(k\) unless several nonzero elements cancel each other.
By identifying the largest zero block at the head of \(\tr^{(j)}\) and the largest zero block at the tail of \(\varrho^{(j)}\) and performing the outer product blockwisely, it's obvious that such cancellations never happen and thus \(Q_j\) equals to the difference between the positions of the last nonzero element in \(\varrho^{(j)}\) and the first nonzero element in \(\tr^{(j)}\).
Since only the upper triangular part of the outer product contributes to the polynomial, if the difference was calculated to be negative or we cannot find a nonzero position for some vectors, the polynomial would be zero (\(Q_j=-\infty\)).
However, since the similarity transformation that transforms \(\mathcal L_{\boldsymbol\chi}\) into its Jordan normal form may be complicated, it might be plausible to expect a low sparsity of these vectors and thus a nonzero \(\mathcal W_j\).


Taking the logarithm of \(G(t)\) in \Cref{eq:G_expression} followed by a time derivative, we obtain \(\partial_t\ln G(t)\) as
\begin{equation}\label{eq:lnG_derivative}
    \dfrac{\sum_{j=1}^\Xi e^{t\lambda_j}\left\llangle\tr^{(j)}\middle|(\lambda_j\mathbb I_j+N_j)e^{tN_j}\middle|\varrho^{(j)}\right\rrangle}{G(t)},  
\end{equation}
where the inner product \(\bbracckett{\tr^{(j)}}{N_je^{tN_j}}{\varrho^{(j)}}\) is a similar polynomial that has one order less than \(\bbracckett{\tr^{(j)}}{e^{tN_j}}{\varrho^{(j)}}\).
When the latter is zero, the former is also zero.

For summands where \(\mathcal W_j=0\), we will omit these zero terms from the summation and exclude them from the range of index symbol \(j\) from now on.
Let \(\Lambda:=\max(\{\operatorname{Re}\lambda_j\}_j)\) denote the maximum real part of the eigenvalues occurring in the remaining sum and decompose the complex eigenvalues \(\lambda_j\triangleq a_j+ib_j\) with \(a_j,b_j\in\mathbb R\).
The numerator of \Cref{eq:lnG_derivative} is asymptotically equivalent to the sum of terms where \(a_j=\Lambda\)
\begin{equation}\label{eq:SCGF_convergence_numerator_asymptotic1}
    e^{t\Lambda}\,\sum_{\mathclap{\substack{j=1\\\text{where}\ a_j=\Lambda}}}^\Xi\, (\Lambda+ib_j)\exp\left(itb_j\right)\mathcal W_j.
\end{equation}
To establish this asymptotic equivalence, we demonstrate that terms with \(a_j<\Lambda\) are asymptotically negligible compared to those with \(a_j=\Lambda\).
Consider the comparison between any term with \(a_j<\Lambda\) and any term with \(a_k=\Lambda\):
\begin{equation}
    \begin{split}
    &e^{ta_j}\bbracckett{\tr^{(j)}}{(\lambda_j\mathbb I_j+N_j)e^{tN_j}}{\varrho^{(j)}}\\
    \ll{}
    &e^{t\Lambda}\bbracckett{\tr^{(k)}}{(\lambda_k\mathbb I_k+N_k)e^{tN_k}}{\varrho^{(k)}},
    \end{split}
\end{equation}
where \(\ll\) denotes the Bachmann-Landau little-o relation.
Since \(e^{itb_j}=O(1)\), we can apply the property \(o(h_1)O(h_2)=o(h_1h_2)\) to obtain \(\text{LHS}\cdot e^{itb_j}\ll\text{RHS}\).
Furthermore, since \(|e^{itb_k}|=1\), the squeeze theorem ensures that \(\text{LHS}\cdot e^{itb_j}\ll\text{RHS}\cdot e^{itb_k}\).
Consequently, all terms with \(a_j<\Lambda\) become asymptotically negligible in the long-time limit, and the numerator of \Cref{eq:lnG_derivative} is asymptotically equivalent to
\begin{equation}\label{eq:SCGF_convergence_numerator_asymptotic2}
    e^{t\Lambda}\,\sum_{\mathclap{\substack{j=1\\\text{where}\ a_j=\Lambda}}}^\Xi\, \exp\left(itb_j\right)\bbracckett{\tr^{(j)}}{(\lambda_j\mathbb I_j+N_j)e^{tN_j}}{\varrho^{(j)}}.
\end{equation}
To complete the derivation, we note that each inner product \(\bbracckett{\tr^{(j)}}{(\lambda_j\mathbb I_j+N_j)e^{tN_j}}{\varrho^{(j)}}\) is asymptotically equivalent to \(\lambda_j\mathcal W_j\) in the long-time limit.
Substituting this asymptotic equivalence into \Cref{eq:SCGF_convergence_numerator_asymptotic2} yields \Cref{eq:SCGF_convergence_numerator_asymptotic1}.
We remark that this substitution preserves the asymptotic equivalence since the multiplicative factor \(e^{t\Lambda}\exp\left(itb_j\right)\) is nonzero.

Invoking the same asymptotic analysis, we can show that the denominator \(G(t)\) in \Cref{eq:lnG_derivative} is asymptotically equivalent to
\begin{equation}\label{eq:SCGF_convergence_denominator_asymptotics}
    e^{t\Lambda}\,\sum_{\mathclap{\substack{j=1\\\text{where}\ a_j=\Lambda}}}^\Xi\, \exp\left(itb_j\right)\mathcal W_j.
\end{equation}
Expanding the factor \(\Lambda+ib_j\) in \Cref{eq:SCGF_convergence_numerator_asymptotic1} into two terms and dividing by the asymptotic equivalent of \(G(t)\) given in \Cref{eq:SCGF_convergence_denominator_asymptotics}, the first term converges to \(\Lambda\).
Therefore \Cref{eq:SCGF_convergence_numerator_asymptotic1} converges if and only if the second term contributed by \(ib_j\) converges.
This second term is
\begin{equation}\label{eq:SCGF_convergence_second_term}
    \left.i\;\sum_{\mathclap{\substack{j=1\\\text{where}\ a_j=\Lambda}}}^\Xi\, \exp\left(itb_j\right)b_j\mathcal W_j\,\middle/\;\sum_{\mathclap{\substack{j=1\\\text{where}\ a_j=\Lambda}}}^\Xi\, \exp\left(itb_j\right)\mathcal W_j.\right.
\end{equation}
Since \(\mathcal W_j\) is a polynomial in \(t\) of order \(Q_j\), only the \(t^{Q_j}\) term contributes to the sum as \(t \to \infty\), leaving only its associated coefficient, which we denote as \(w_j\).

At this point, both the numerator and denominator of the term in \Cref{eq:SCGF_convergence_second_term} are trigonometric polynomials.
For this fraction to converge, all occurring \(b_j\) values must be identical (say, equal to \(b\)) such that \Cref{eq:SCGF_convergence_second_term} approaches \(ib\).
This implies that among the eigenvalues with the maximum real part \(\Lambda\), only one distinct imaginary part is allowed.
This can be demonstrated as follows: Assume that this term converges to \(i\mathcal B\), i.e. (summation range omitted),
\begin{equation}
    \lim_{t\to\infty}\frac{\sum_j \exp\left(itb_j\right)(b_j-\mathcal B)w_j}{\sum_j \exp\left(itb_j\right)w_j}=0
\end{equation}
Since its denominator is bounded, its numerator must converge to \(0\) as \(t\to\infty\).
Thus the average power density (over a period) of this trigonometric polynomial, which is a constant equaling to the Euclidean norm of polynomial coefficients (by Parseval's theorem), must also be \(0\).
This necessitates that all coefficients \(b_j-\mathcal B\) of the trigonometric polynomial are \(0\).
Obviously this condition is also sufficient.

For an example where a convergence is not achievable, we consider an extreme case where a complex conjugate pair \(\Lambda\pm ib\) share the same maximum real part \(\Lambda\). The fraction in \Cref{eq:SCGF_convergence_second_term} thus becomes
\begin{equation}
    \begin{split}
    &i\frac{\exp\left(itb\right)b\varrho_+-\exp\left(-itb\right)b\varrho_-}{\exp\left(itb\right)\varrho_++\exp\left(-itb\right)\varrho_-}\\
    ={}&
    ib\tanh\left[\frac{1}{2}\ln\left(\frac{\varrho_+}{\varrho_-}\right)+itb\right],
    \end{split}
\end{equation}
where \(\varrho_\pm\) are diagonal elements of \(\varrho\) and correspond to the \(1\times1\) Jordan blocks associated with \(\Lambda\pm ib\). In the case of \(\varrho_+=\varrho_-\), the above function reduces to a real-valued unbounded periodic function \(-b\tan(bt)\).

In summary, the general procedure for calculating the limit \(\displaystyle\lim_{t\to\infty}\partial_t\ln G(t)\) is as follows:

\begin{enumerate}
    \item After vectorizing \(\varrho,\tr\) and \(\mathcal L_{\boldsymbol\chi}\), switching to a basis where \(\mathcal L_{\boldsymbol\chi}\) is in its Jordan canonical form.
    \item Partitioning the vectors \(\tr,\varrho\) into segments \(\tr^{(j)},\varrho^{(j)}\) according to the Jordan blocks. Calculating the polynomial order \(Q_j\) of \(\mathcal W_j\) by subtracting the index of the first nonzero element in \(\tr^{(j)}\) and the index of the last nonzero element in \(\varrho^{(j)}\). Discarding any blocks where this difference is negative or one of these two vectors is zero.
    \item From the remaining Jordan blocks, identifying the eigenvalues with the largest real part \(\Lambda\) and discarding all other blocks.
    \item Determining the largest \(Q_j\) among the remaining blocks. Discarding any blocks with a lower order. For the diagonalizable case, all block sizes are \(1\times1\). Therefore the highest order is \(0\) and no block should be further discarded.
    \item If the eigenvalues in remaining blocks have different imaginary parts, the limit does not converge.
    \item If all remaining blocks correspond to the same eigenvalue (that is, they have the same real part \(\Lambda\) and the same imaginary part \(b\)), then the limit converges to this eigenvalue \(\Lambda + ib\).
\end{enumerate}

It is common to compute the SCGF simply as the eigenvalue with the largest real part.
However, the condition shown above hints at the possibility of constructing counterexamples where the SCGF does not simply converge to the eigenvalue with the largest real part. When selecting eigenvalues, tolerance should also be considered to avoid incorrect selection due to numerical errors.

\renewcommand{\theequation}{B\arabic{equation}}
\setcounter{equation}{0}  
\section{The condition to ensure the fluctuation symmetry}\crefalias{section}{appendix}\label{sec:condition_for_fluctuation_theorem}

\subsection{Overview}
In this appendix, we derive the necessary and sufficient conditions for the vectorized tilted Liouvillian superoperator to satisfy the relation
\begin{equation}\label{eq:symmetry_liouville_even}
	\mathcal L_{\boldsymbol\chi}^T=\mathcal L_{-\boldsymbol\chi+i\boldsymbol\sigma}.
\end{equation}
Here, \(\boldsymbol\sigma\) is a vector of thermodynamic affinities. Since the transpose operation preserves the eigenvalue spectrum, this relation directly implies the fluctuation symmetry of SCGF given in \Cref{eq:symmetry_scgf} of the main text. To proceed, we assume the existence of a basis where both the Hamiltonian and the jump operators are represented by real matrices. This assumption is naturally satisfied in our model when using the tensor product basis formed from number states of electrons and photons, and is closely related to certain time-reversal symmetry~\cite{sakurai_symmetry_2020,landi_current_2024}.

\subsection{Necessary and sufficient conditions}
We begin with the complete form of the vectorized tilted Liouvillian involving $m$ channels
\begin{equation}\label{eq:vectorized_liouvillian}
    \begin{split}
	\mathcal L_{\boldsymbol\chi}
    =
    {}-{}&i(I\otimes H-H^T\otimes I)\\
    {}+{}&\sum_{j=1}^me^{i\chi_{\alpha(\nu_j)}}L_j^*\otimes L_j\\
    {}-{}&\sum_{j=1}^m\cfrac{I\otimes L_j^\dagger L_j+(L_j^\dagger L_j)^T\otimes I}{2},
    \end{split}
\end{equation}
where \(I\) is an identity matrix. The counting fields may depend on the index $j$ of the jump operators; for brevity, we suppress this dependency in the notation. This formulation corresponds to counting weighted jump events, and such random variables remain consistent with the characteristic functions defined earlier, in accordance with the projection-slice theorem\cite{deans_radon_1983,Ng.05.TOG}. It is important to note that the jump counts themselves do not inherently satisfy the symmetry expressed in \Cref{eq:symmetry_liouville_even}. To facilitate further analysis, we fix the usage of several index symbols; their purposes, ranges, and mutual dependencies are summarized in \Cref{tab:index_symbols}.

\begin{table*}[htbp]
\caption{Index symbols and their ranges, purposes and mutual dependencies. The admissible counting field indices mentioned in the text are identical to possible current numbers, e.g. in \Cref{tab:channel_numbering} for a specific model.\label{tab:index_symbols}}
	\small
	\begin{tabular}{|c|c|c|c|}
		\hline
		Index & Range & Purpose & Depends on \\
		\hline
		\(j\) & \(1,2,\cdots,m\) or smaller & mark different jump channels in the sum & none \\
		\hline
		\(\alpha\) & all counting field indices & mark the counting field for \(j\)-th channel & \(j\) \\
		\hline
		\(\mu\) & all counting field indices & \parbox{340pt}{mark counting field dimensions that trivially transform to \(\delta(n_\mu)\) in multi-dimensional Fourier transform, except the nontrivial dimension \(\alpha\) for current \(j\)-th term} & \(j\) \\
		\hline
		\(\iota\) & all counting field indices & a free variable marking the selected counting field, which is summed over later & none \\
		\hline
		\(\eta\) & \(1,2,\dots,\operatorname{Card}(Z_\iota)\) & mark distinct positive counting weights corresponding to \(\chi_\iota\) without repetition & \(\iota\) \\
        \hline
	\end{tabular}
\end{table*}

Under the assumption of real matrices (and self-adjoint Hamiltonian), \Cref{eq:vectorized_liouvillian} simplifies to
\begin{equation}\label{eq:vectorized_liouvillian_real_basis}
    \begin{split}
	\mathcal L_{\boldsymbol\chi}
    =
    {}-{}&i(I\otimes H-H^T\otimes I)\\
    {}+{}&\sum_{j=1}^me^{i\chi_\alpha\nu_j}L_j\otimes L_j\\
    {}-{}&\sum_{j=1}^m\cfrac{I\otimes L_j^T L_j+L_j^T L_j\otimes I}{2}.
    \end{split}
\end{equation}
The matrices in \Cref{eq:symmetry_liouville_even} then read
\begin{equation}\label{eq:b4}
    \begin{split}
	\mathcal L_{\boldsymbol\chi}^T
    =
    {}-{}&i(I\otimes H-H^T\otimes I)\\
    {}+{}&\sum_{j=1}^me^{i\chi_\alpha\nu_j}L_j^T\otimes L_j^T\\
    {}-{}&\sum_{j=1}^m\cfrac{I\otimes L_j^T L_j+L_j^T L_j\otimes I}{2}.
    \end{split}
\end{equation}
\begin{equation}\label{eq:b5}
    \begin{split}
	\mathcal L_{-\boldsymbol\chi+i\boldsymbol\sigma}
    =
    {}-{}&i(I\otimes H-H^T\otimes I)\\
    {}+{}&\sum_{j=1}^me^{-i\chi_\alpha\nu_j}e^{-\nu_j\sigma_\alpha}L_j\otimes L_j\\
    {}-{}&\sum_{j=1}^m\cfrac{I\otimes L_j^T L_j+L_j^T L_j\otimes I}{2}.
    \end{split}
\end{equation}
Inserting \Cref{eq:b4,eq:b5} into \Cref{eq:symmetry_liouville_even} and collecting all \(\boldsymbol\chi\)-dependent terms on the left and the \(\boldsymbol\chi\)-independent terms on the right, we obtain the following equivalent form 
\begin{equation}\label{eq:b6}
	\sum_{j=1}^me^{i\chi_\alpha\nu_j}L_j^T\otimes L_j^T-\sum_{j=1}^me^{-i\chi_\alpha\nu_j}e^{-\nu_j\sigma_\alpha}L_j\otimes L_j=0.
\end{equation}
We then look for conditions that guarantee the validity of the above equation. Since \Cref{eq:b6} should hold for any \(\boldsymbol\chi\), we can regard \(\boldsymbol\chi\) as the frequency domain variable and perform the inverse Fourier transform of \Cref{eq:b6}, yielding
\begin{equation}\label{eq:dist_from_fourier}
    \begin{split}
    &\sum_{j=1}^mL_j^T\otimes L_j^T\delta\left(n_\alpha-\nu_j\right)\prod_{\mu\neq\alpha}\delta\left(n_\mu\right)\\
    ={}&\sum_{j=1}^me^{-\nu_j\sigma_\alpha}L_j\otimes L_j\delta\left(n_\alpha+\nu_j\right)\prod_{\mu\neq\alpha}\delta\left(n_\mu\right).
    \end{split}
\end{equation}
For these two linear combinations of Dirac delta distributions to be equal, the coefficients at each distinct peak position must be equal.
For simplicity, we assume that the indices \(j\) are ordered such that all terms with \(\nu_j=0\) are placed at the end and the last nonzero weight appears at position \(m'\).
The equality of coefficients at \(\boldsymbol n=\boldsymbol 0\) then yields the following condition for the zero-weight terms:
\begin{equation}\label{eq:zero_weight_condition_even}
	\sum_{j=m'+1}^{m}L_j^T\otimes L_j^T
	=\sum_{j=m'+1}^{m}L_j\otimes L_j.
\end{equation}
Technically, obtaining \Cref{eq:zero_weight_condition_even} from \Cref{eq:dist_from_fourier} is performed by constructing a test function \(\phi_0\) that vanishes everywhere except in a sufficiently small neighborhood of \(\boldsymbol 0\) and applying the linear combinations of Dirac delta distributions as a linear functional to \(\phi_0\):
\begin{equation}
    \begin{aligned}
        &\bigintssss \left( \sum_{j=1}^mL_j^T\otimes L_j^T\delta\left(n_\alpha-\nu_j\right)\prod_{\mu\neq\alpha}\delta\left(n_\mu\right) \right) \phi_0(\boldsymbol n) \dd \boldsymbol n\\
        =&
        \bigintssss \left( \sum_{j=1}^me^{-\nu_j\sigma_\alpha}L_j\otimes L_j\delta\left(n_\alpha+\nu_j\right)\prod_{\mu\neq\alpha}\delta\left(n_\mu\right) \right) \phi_0(\boldsymbol n) \dd \boldsymbol n.
    \end{aligned}
\end{equation}
Such test functions are known as bump functions; their existence and construction methods can be found in \cite{nlab:bump_function,lee_manifolds_2012}.

Using \Cref{eq:zero_weight_condition_even}, the terms with \(\nu_j=0\) in \Cref{eq:dist_from_fourier} are canceled out, yielding
\begin{equation}
    \begin{split}
	&\sum_{j=1}^{m'}L_j^T\otimes L_j^T\delta\left(n_\alpha-\nu_j\right)\prod_{\mu\neq\alpha}\delta\left(n_\mu\right)\\
    ={}&\sum_{j=1}^{m'}e^{-\nu_j\sigma_\alpha}L_j\otimes L_j\delta\left(n_\alpha+\nu_j\right)\prod_{\mu\neq\alpha}\delta\left(n_\mu\right).
    \end{split}
\end{equation}
Using a shorthand of multi-dimensional Dirac delta \(\boldsymbol\delta\) and denoting \(\alpha\)-th standard basis vector as \(\boldsymbol e_\alpha:=(0,\dots,1,\dots,0)\), we can rewrite the above equation as
\begin{equation}\label{eq:other_weight_condition_multidim_delta}
    \sum_{j=1}^{m'}L_j^T\otimes L_j^T\boldsymbol\delta\left(\boldsymbol n-\nu_j\boldsymbol e_\alpha\right)=
    \sum_{j=1}^{m'}e^{-\nu_j\sigma_\alpha}L_j\otimes L_j\boldsymbol\delta\left(\boldsymbol n+\nu_j\boldsymbol e_\alpha\right).
\end{equation}
In the above equality, the \(j\)-th term in the two summations contributes a multi-dimensional Dirac delta located on the \(\alpha\)-th axis at coordinate \(\pm\nu_j\) (\(+\nu_j\) for the left-hand side and \(-\nu_j\) for the right-hand side), i.e., located at \((0,\dots,\underbrace{\pm\nu_j}_{\mathclap{\alpha\text{-th}}},\dots,0)\). If the left-hand side of \Cref{eq:other_weight_condition_multidim_delta} has a nonvanishing term at \(n_\alpha=\nu_j\), then its right-hand side must have a nonvanishing term at \(n_\alpha=-\nu_j\) for this equality to hold.
Therefore, we must require the set of all weights \(\nu_j\) associated with a fixed counting field \(\chi_\alpha\) is symmetric about zero, i.e. it can be written as \(Z_\alpha\cup -Z_\alpha\); Here, \(Z_\alpha\) is the subset formed by all positive weights associated with \(\chi_\alpha\).

Since there may be multiple terms indexed by different \(j\) sharing the same peak location, we define a new index \(\eta\) to label distinct positive weights \(\nu_\eta\) in \(Z_\alpha\) without repetition and collect terms with the same \(\nu_j>0\) (\(\nu_j<0\)) on the left-hand side (right-hand side) of \Cref{eq:other_weight_condition_multidim_delta}:
\begin{equation}\label{eq:other_weight_condition_even_multidim_delta_presummation}
    \begin{split}
	&\sum_\iota\;\sum_{\mathclap{\nu_\eta\in Z_\iota}}\;\boldsymbol\delta\left(\boldsymbol n-\nu_\eta\boldsymbol e_\iota\right)\,\sum_{\mathclap{\substack{j=1\\\text{where}\ \alpha=\iota\\\text{and}\ \nu_j=\nu_\eta}}}^{m'}\,L_j^T\otimes L_j^T\\
    {}+{}&\text{terms where }\nu_j<0\\
    {}={}&\sum_\iota\;\sum_{\mathclap{\nu_\eta\in Z_\iota}}\;\boldsymbol\delta\left(\boldsymbol n-\nu_\eta\boldsymbol e_\iota\right)e^{\nu_\eta\sigma_\iota}\,\sum_{\mathclap{\substack{j=1\\\text{where}\ \alpha=\iota\\\text{and}\ \nu_j=-\nu_\eta}}}^{m'}\,L_j\otimes L_j\\
    {}+{}&\text{terms where }\nu_j>0.
    \end{split}
\end{equation}
We can obtain an equation for the coefficients of each \(\boldsymbol\delta\left(\boldsymbol n-\nu_\eta\boldsymbol e_\iota\right)\) by constructing a bump function supported in small neighborhoods of each peak location:
\begin{equation}\label{eq:other_weight_condition_even}
	\sum_{\mathclap{\substack{j=1\\\text{where}\ \alpha=\iota\\\text{and}\ \nu_j=\nu_\eta}}}^{m'}L_j^T\otimes L_j^T
    \;=\;
    e^{\nu_\eta\sigma_\iota}\sum_{\mathclap{\substack{j=1\\\text{where}\,\alpha=\iota\\\text{and}\ \nu_j=-\nu_\eta}}}^{m'}\,L_j\otimes L_j.
\end{equation}

For the coefficient-wise equation obtained from the omitted terms in \Cref{eq:other_weight_condition_even_multidim_delta_presummation}, it is easy to see that these equations are simply duplicates of \Cref{eq:other_weight_condition_even}, obtained by moving the \(e^{\nu_\eta\sigma_\iota}\) to the left side and taking the transpose of both sides.

In addition, it is worth noting that if we are simply counting the jump events themselves, the range of counting weights for one counting field contains only one element. As a corollary of requiring the weights to be symmetric around zero, each weight must be zero. Thus, no actual counting is present and the condition for the fluctuation theorem reduces to \Cref{eq:zero_weight_condition_even}, which in turn ensures the stronger time reversibility~\cite{strasberg_thermodynamics_2022} for quantum Markovian dynamics.

In summary, under the real-matrix assumption, the conjunction of Eqs. \eqref{eq:zero_weight_condition_even} and \eqref{eq:other_weight_condition_even} form necessary and sufficient conditions for the SCGF to satisfy the fluctuation symmetry given in \Cref{eq:symmetry_scgf} of the main text. These conditions also enforce that the set of admissible weights \(\nu\) is symmetric around zero. For clarity, we summarize the procedure of the proof as follows:
\begin{enumerate}
    \item Separate symmetry relation into two parts where zero and nonzero counting weights are considered separately.
    \item Perform multi-dimensional Fourier transform to obtain equations of distributions. The resulting distributions are linear combinations of Dirac delta distributions \(\boldsymbol\delta\left(\boldsymbol n-\nu_\eta\boldsymbol e_\iota\right)\).
    \item Apply distributions to bump functions to isolate the equality for each distinct \(\nu_\eta\boldsymbol e_\iota\). This will be equations for the coefficients of the previously mentioned linear combination.
\end{enumerate}

\subsection{Solely sufficient condition}
Both \Cref{eq:zero_weight_condition_even} and \Cref{eq:other_weight_condition_even} equate linear combinations of tensor products.
The equality of two sums (denoted \(\sum_j\hat A_j\) and \(\sum_j\hat B_j\)) does not, in general, imply pairwise equality of the summands (i.e., \(\{\hat A_j\}_j=\{\hat B_j\}_j\)).
While pairwise equality would guarantee equality of the sums, one may arbitrarily insert a cancelling pair \(\hat K\otimes\hat K\) and \(-\hat K\otimes\hat K\) on either side, keeping the equation valid even though the number of terms differs.
Even if we limit the sums to be composed of a fixed number of linearly independent operators, it is still possible to replace two operators \(\hat C\otimes\hat C\) and \(\hat D\otimes\hat D\) by, for instance,
\((\hat C\cos\varphi\pm\hat D\sin\varphi)\) and \((\hat C\sin\varphi\mp\hat D\cos\varphi)\), with arbitrary \(\varphi\in\mathbb R\).
In this way the number of terms and linear independence can be preserved while individual terms change, so pairwise equality need not hold.

A simple case is when either \(\{\hat A_j\}_j\) or \(\{\hat B_j\}_j\) contains only one operator, with the former being \(\hat L_+^T\otimes\hat L_+^T\) and the latter \(e^{-\sigma_\iota\nu_\eta}\hat L_-\otimes\hat L_-\).
This reduces to
\begin{equation}\label{eq:single_operator_condition}
    \hat L_+^T\otimes\hat L_+^T=e^{\sigma_\iota\nu_\eta}\hat L_-\otimes\hat L_-,
\end{equation}
where the only two nontrivial solutions are
\begin{equation}\label{eq:single_operator_condition_decomposed}
    \hat L_+^T=\pm e^{\sigma_\iota\nu_\eta/2}\hat L_-,
\end{equation}
which mean that for each jump operator \(L_j\) with \(\nu_j=\nu_\eta\), there must exist another jump operator \(\pm e^{-\sigma_\iota\nu_\eta/2}L_j^T\) with \(\nu_j=-\nu_\eta\).
This is often referred to as a detailed balance condition~\cite{landi_current_2024}
\begin{equation}\label{eq:detailed_balance}
    \hat L_+=e^{\sigma_\iota\nu_\eta/2}\hat L_-^\dagger,
\end{equation}
since transposing in the real-matrix representation is equivalent to Hermitian adjoint and we lose no information about Lindblad equations while fixing the above sign of a jump operator.

\subsection{Proof of tensor product factorization}

The statement used in reducing \Cref{eq:single_operator_condition} to \Cref{eq:single_operator_condition_decomposed} is an immediate corollary in finite-dimensional matrix representations of the following proposition.
\begin{proposition}
	For any linear operators \(X,Y\), if \(X\otimes X=Y\otimes Y\), then \(X=\pm Y\). (Here \(\otimes\) represents the tensor product.)
\end{proposition}

If \(X=0\) or \(Y=0\) the claim is obvious.
For the nonzero case, it suffices to prove that for any nonzero \(A,B,C,D\),
\begin{equation}\label{eq:general_tensor_product_factorization}
	A\otimes B=C\otimes D\implies \exists c\in\mathbb F,\quad A=c C,D=c B,
\end{equation}
where \(\mathbb F\) is the field of the underlying vector space.
Setting \(A=B=X\) and \(C=D=Y\) then yields \(X=c^2X\), hence \(c=\pm1\).

Since \(A\otimes B\) and \(C\otimes D\) act identically on all vectors, for arbitrary \(\ket{a}\otimes\ket{b}\) we have
\begin{equation}\label{eq:tensor_cartesian_on_vector}
	A\ket{a}\otimes B\ket{b}=C\ket{a}\otimes D\ket{b}.
\end{equation}
Since the tensor product of two vectors is unique up to a scalar multiple, i.e., for any nonzero \(\ket u,\ket u',\ket v,\ket v'\)
\begin{equation}\label{eq:tensor_cartesian}
    \begin{split}
	&\ket{u}\otimes \ket{v}=\ket{u'}\otimes \ket{v'}\\
    \implies\quad
    &\exists c\in\mathbb F,\quad\ket{u}=c \ket{u'}, \ket{v'}=c\ket{v},
    \end{split}
\end{equation}
we can choose \(\ket{a_1}\notin\ker A\) and \(\ket{b_1}\notin\ker B\) to obtain
\begin{equation}
	A\ket{a_1}=cC\ket{a_1}\ \land\ D\ket{b_1}=cB\ket{b_1}
\end{equation}
for some \(c\). Substituting \(\ket{a_1}\) into \Cref{eq:tensor_cartesian_on_vector} yields
\begin{equation}
	C\ket{a_1}\otimes (D-cB)\ket{b}=0.
\end{equation}
Since the left part of the tensor product is nonzero, we must have \((D-cB)\ket{b}=0\), and since \(\ket b\) is arbitrary, we have \(D=cB\) along with \(c\) being uniquely determined.
Similarly, substituting \(\ket{b_1}\) back to \Cref{eq:tensor_cartesian_on_vector} gives \(A=cC\).

Property \Cref{eq:tensor_cartesian} used above follows from the universal property of the tensor product.
Let \(V,W\) be the underlying vector spaces.
For linear functionals \(\bra\phi\in V^*\), \(\bra\psi\in W^*\), the bilinear map \((\ket u,\ket v)\mapsto\braket{\phi|u}\braket{\psi|v}\) induces a linear functional on \(V\otimes W\).
Applying it to the premise of \Cref{eq:tensor_cartesian} yields
\begin{equation}\label{eq:bilinear_on_cartesian}
	\braket{\phi|u}\braket{\psi|v}=\braket{\phi|u'}\braket{\psi|v'}.
\end{equation}
Choose \(\bra{\psi_0}\) with \(\braket{\psi_0|v}\neq0\). Then \(\braket{\phi|u}=c\braket{\phi|u'}\) with \(c=\braket{\psi_0|v'}/\braket{\psi_0|v}\).
Since the only vector mapped to zero by all linear functionals is the zero vector, we deduce \(\ket u-c\ket{u'}=0\), and \(c\) can be uniquely determined by \(\ket u\) and \(\ket{u'}\).
Substituting \(\ket u=c\ket{u'}\) back to \Cref{eq:bilinear_on_cartesian} and using the arbitrariness of \(\bra\phi\) again gives \(\ket v'=c\ket v\).

%

\end{document}